\documentclass[twocolumn]{aastex63}
\usepackage{amsmath}
\usepackage{mathtools}  
\usepackage{comment}
\usepackage{graphicx}
\usepackage{xcolor}
\usepackage{xparse,xcoffins}
\usepackage{appendix}

\shorttitle{$\sigma(v)$ mapping}
\shortauthors{Yang et al.}

\ExplSyntaxOn
\NewCoffin\imagecoffin
\NewCoffin\labelcoffin

\keys_define:nn { miguel/label }
 {
  label   .tl_set:N = \l_miguel_label_tl,
  labelbox .bool_set:N = \l_miguel_label_box_bool,
  labelbox .default:n = true,
  fontsize .tl_set:N = \l_miguel_label_size_tl,
  fontsize .initial:n = \footnotesize,
  pos .choice:,
  pos/nw .code:n = \tl_set:Nn \l_miguel_label_pos_tl { left,up },
  pos/ne .code:n = \tl_set:Nn \l_miguel_label_pos_tl { right,up },
  pos/sw .code:n = \tl_set:Nn \l_miguel_label_pos_tl { left,down },
  pos/se .code:n = \tl_set:Nn \l_miguel_label_pos_tl { right,down },
  pos/n .code:n = \tl_set:Nn \l_miguel_label_pos_tl { hc,up },
  pos/w .code:n = \tl_set:Nn \l_miguel_label_pos_tl { left,vc },
  pos/s .code:n = \tl_set:Nn \l_miguel_label_pos_tl { hc,down },
  pos/e .code:n = \tl_set:Nn \l_miguel_label_pos_tl { right,vc },
  pos .initial:n = nw,
  unknown .code:n   = \clist_put_right:Nx \l_miguel_label_clist
                       { \l_keys_key_tl = \exp_not:n { #1 } }
 }
\clist_new:N \l_miguel_label_clist
\box_new:N \l_miguel_label_box
\box_new:N \l_miguel_label_image_box

\NewDocumentCommand{\xincludegraphics}{O{}m}
 {
  \group_begin:
  \tl_clear:N \l_miguel_label_tl
  \clist_clear:N \l_miguel_label_clist
  \keys_set:nn { miguel/label } { #1 }
  \tl_if_empty:NTF \l_miguel_label_tl
   {
    \miguel_includegraphics:Vn \l_miguel_label_clist { #2 }
   }
   {
    \SetHorizontalCoffin\imagecoffin
     {
      \miguel_includegraphics:Vn \l_miguel_label_clist { #2 }
     }
    \SetHorizontalCoffin\labelcoffin
     {
      \raisebox{\depth}
       {
        \bool_if:NTF \l_miguel_label_box_bool
         { \fcolorbox{white}{white}{\l_miguel_label_size_tl\l_miguel_label_tl} }
         { \l_miguel_label_size_tl\l_miguel_label_tl }
       }
     }
    \SetVerticalPole\imagecoffin{left}{3pt+\CoffinWidth\labelcoffin/2}
    \SetVerticalPole\imagecoffin{right}{\Width-3pt-\CoffinWidth\labelcoffin/2}
    \SetHorizontalPole\imagecoffin{up}{\Height-3pt-\CoffinHeight\labelcoffin/2}
    \SetHorizontalPole\imagecoffin{down}{3pt+\CoffinHeight\labelcoffin/2}
    \use:x{\JoinCoffins\imagecoffin[\l_miguel_label_pos_tl]\labelcoffin[vc,hc]} 
    \TypesetCoffin\imagecoffin
   }
   \group_end:
 }
\NewDocumentCommand{\setlabel}{m}
 {
  \keys_set:nn { miguel/label } { #1 }
 }

\cs_new_protected:Nn \miguel_includegraphics:nn
 {
  \includegraphics[#1]{#2}
 }
\cs_generate_variant:Nn \miguel_includegraphics:nn { V }

\ExplSyntaxOff

\begin{document}

\title{Gravothermal solutions of SIDM halos: mapping from constant to velocity-dependent cross section}

\correspondingauthor{Shengqi Yang}
\email{syang@carnegiescience.edu}

\author{Shengqi Yang}
\affiliation{Carnegie Observatories, 813 Santa Barbara Street, Pasadena, CA 91101, U.S.A}

\author{Xiaolong Du}
\affiliation{Carnegie Observatories, 813 Santa Barbara Street, Pasadena, CA 91101, U.S.A}

\author{Zhichao Carton Zeng}
\affiliation{Department of Physics, The Ohio State University, 191 W. Woodruff Ave., Columbus, OH 43210, USA}
\affiliation{Center for Cosmology and AstroParticle Physics, The Ohio State University, 191 W. Woodruff Ave., Columbus, OH 43210, USA}

\author{Andrew Benson}
\affiliation{Carnegie Observatories, 813 Santa Barbara Street, Pasadena, CA 91101, U.S.A}

\author{Fangzhou Jiang}
\affiliation{Carnegie Observatories, 813 Santa Barbara Street, Pasadena, CA 91101, U.S.A}

\author{Ethan O.~Nadler}
\affiliation{Carnegie Observatories, 813 Santa Barbara Street, Pasadena, CA 91101, USA}
\affiliation{Department of Physics $\&$ Astronomy, University of Southern California, Los Angeles, CA, 90007, USA}

\author{Annika H.~G.~Peter}
\affiliation{Department of Physics, The Ohio State University, 191 W. Woodruff Ave., Columbus, OH 43210, USA}
\affiliation{Department of Astronomy, The Ohio State University, 140 W. 18th Ave., Columbus, OH 43210, USA}
\affiliation{Center for Cosmology and AstroParticle Physics, The Ohio State University, 191 W. Woodruff Ave., Columbus, OH 43210, USA}



\begin{abstract}
The scale-free gravothermal fluid formalism has long proved effective in describing the evolution of self-interacting dark matter halos with a constant dark matter particle cross section. However, whether the gravothermal fluid solutions match numerical simulations for velocity-dependent cross- section scenarios remains untested. In this work, we provide a fast mapping method that relates the constant-cross-section gravothermal solution to models with arbitrary velocity dependence in the cross section. We show that the gravothermal solutions after mapping are in good agreement with \textsc{Arepo} N-Body simulation results. We illustrate the power of this approach by applying this fast mapping method to a halo hosting a low surface brightness galaxy UGC 128. We show that this fast mapping method can be used to constrain free parameters in a physically motivated cross-section model and illustrate parameter space favored by the rotation curve measurement.
\end{abstract}

\section{Introduction}
Self-interacting dark matter (SIDM) is a promising class of candidate to explain the cored dark matter (DM) halos implied by observations of local dwarf galaxies \citep{2008AJ....136.2648D,2015AJ....149..180O}, while maintaining the accurate reproduction of large-scale cosmic properties provided by the cold dark matter (CDM) paradigm \citep{1982ApJ...263L...1P,2011A&A...536A...1P,2014A&A...571A..16P,2012MNRAS.427.3435A}. In the SIDM model, first proposed by \cite{2000PhRvL..84.3760S}, in addition to the gravitational interactions, DM particles also scatter with each other. The effect of scattering is negligible on large scales due to the low DM particle number density and the resulting long relaxation time. However, the DM particle collisional mean free path can become smaller than the orbital Jeans scale near the halo center, causing the formation of an isothermal core and even the gravitational catastrophe \citep{1968MNRAS.138..495L} at small scales.\par 
The gravothermal fluid formalism originally developed for studying the gravitational catastrophe of stellar clusters \citep{1968MNRAS.138..495L,1971MNRAS.151..351E} has been extended to isolated SIDM halos and provides rich physical insights into the nature of SIDM halo evolution (e.g. \citealt{1980MNRAS.191..483L,2002PhRvL..88j1301B,2002ApJ...568..475B,2018PhRvD..98b3021S,Essig2019,2020PhRvD.101f3009N}). Although it makes multiple simplifying assumptions, numerical solutions to the gravothermal fluid formalism show surprisingly good agreement with idealized N-Body simulations and also match reasonably well with cosmological SIDM zoom-in simulations \citep{2011MNRAS.415.1125K,2015MNRAS.453...29E,Essig2019,2022arXiv220406568O}. A nice property of the gravothermal fluid formalism is that the fluid equations themselves can be re-written into a form that is independent of the halo density and radius scales. Therefore, although computationally expensive, the fluid formalism only needs to be numerically solved once and can be rescaled to SIDM halos with arbitrary masses, sizes, and concentrations.\par
All previous comparisons between the gravothermal solutions and numerical simulations have assumed a constant DM particle scattering cross section. However, if SIDM is the explanation for the apparent deviations from CDM behavior on small scales, then recent observations indicate that the cross section, $\sigma$, for DM particle self-interaction is likely varying among systems of different scales. Specifically, $\sigma$ is constrained to be of the order of $0.1$ cm$^2$/g, $1$ cm$^2$/g, and 10 cm$^2$/g for galaxy cluster, galactic, and dwarf-galaxy halos, respectively, where the DM particle velocity dispersion decreases from $\sim$1000 km/s to $\sim$1 km/s (e.g. \citealt{2016PhRvL.116d1302K,2018ApJ...853..109E,2021JCAP...01..024S,2022MNRAS.510...54A}). A velocity-dependent cross section $\sigma\propto v^{-4}$ at high velocity is also motivated by particle physics, assuming a Yukawa or Coulomb potential for the self-interaction \citep{2009PhRvD..79b3519A,2009JCAP...07..004F,2010PhLB..692...70I,2018MNRAS.474..388K}. Motivated by this, more recent SIDM numerical simulations have started to focus on velocity-dependent cross section models (e.g. \citealt{2020ApJ...896..112N}). Testing whether the gravothermal fluid solutions are still in good agreement with numerical simulations assuming a velocity-dependent cross section model becomes crucial in developing better understanding of the SIDM N-Body simulations, as well as justifying SIDM cross section constraints derived by gravothermal fluid formalism-based methods.\par
In this work, we show using an analytic gravothermal fluid formalism that assuming different cross section values/models effectively re-scales SIDM halo evolution rates. In particular, there exists a one-on-one time mapping between any two SIDM halos simulated by the gravothermal fluid approach such that their density, velocity, and all the other physical properties are identical up to some re-scaling factors. Based on this finding, we propose a fast mapping method that transfers the gravothermal numerical solutions assuming a constant SIDM cross section to those with arbitrary velocity dependent cross section models. To test the performance of this mapping method we perform multiple idealized N-Body simulations assuming velocity-dependent cross section scenarios with \textsc{Arepo} \citep{2010MNRAS.401..791S}. We find excellent agreement between the mapped gravothermal solutions and N-Body simulation results. This mapping approach is powerful because it eliminates the degrees of freedom introduced by SIDM cross section model assumptions, such that one only needs to solve the gravothermal fluid equations for once and one can then re-scale the solution to halos of arbitrary sizes, masses, and cross section models. Due to the high computational efficiency, this method is useful for exploring the continuous parameter space and constraining SIDM cross sections. It is also applicable to semi-analytic galaxy formation models to efficiently generate SIDM halo populations. To demonstrate the utility of this approach, we combine the mapping method with rotation curve measurements of the low surface brightness (LSB) galaxy UGC 128 \citep{1993AJ....106..548V} and illustrate the parameters of the Navarro–Frenk–White (NFW) density profile \citep{1996ApJ...462..563N} and velocity-dependent cross section models favored by observations.\par
As this work was nearing completion, \citeauthor{2022arXiv220406568O}~(\citeyear{2022arXiv220406568O}; hereafter \defcitealias{2022arXiv220406568O}{O22}\citetalias{2022arXiv220406568O}) introduced a mapping method with a similar spirit, but which differs in detail from our work. We find that \citetalias{2022arXiv220406568O} only compared the mapped gravothermal solutions with idealized N-Body simulations that assume constant cross sections \citep{2011MNRAS.415.1125K}. Example applications of their mapping method are also not discussed in detail. We have therefore expanded this work to include comparison of our approach with that of \citetalias{2022arXiv220406568O}.\par
Calibrating the gravothermal fluid formalism to cosmological zoom-in simulations is an even more interesting topic to explore. However, cosmological simulations generally contains more complicated physics and can break multiple assumptions made in the current fluid model. As a first step, in this work we will only compare gravothermal solutions with idealized SIDM N-Body simulations for isolated halos with constant or velocity dependent cross sections. We leave the more complicated gravothermal solution versus cosmological simulation comparisons to future works.\par 
The plan of this paper is as follows. In Section~\ref{sec:Arepo} we briefly introduce our idealized SIDM halo simulation sets. Section~\ref{sec:gravoIntro} is an introduction to the gravothermal fluid formalism. We briefly review the long-mean-free-path (lmfp) gravothermal solution mapping method among different constant cross scenarios introduced by \cite{2002ApJ...568..475B,2011MNRAS.415.1125K,Essig2019} in Section~\ref{sec:mappingEssig}. We then explain the methodology of mapping the lmfp gravothermal solutions from constant cross section $\sigma$ to velocity dependent cross section $\sigma(v)$ in Section~\ref{sec:mapping}, and comparing the mapped gravothermal solutions with N-Body simulation results as well as \citetalias{2022arXiv220406568O} in Section~\ref{sec:matchArepo}. We briefly discuss the generalization of this lmfp mapping method to the short-mean-free-path (smfp) and intermediate regime in Section~\ref{sec:generalization}. In Section~\ref{sec:pConstrain} we combine the mapping method with observational data and show parameter constraints of the velocity-dependent cross section models. We conclude in Section~\ref{sec:concluson}.
\section{Idealized SIDM N-Body simulation}\label{sec:Arepo}
In this work we perform idealized N-Body simulations of an isolated Milky Way-sized halo with NFW initial density profile:
\begin{equation}\label{eq:NFW}
    \rho(r)=\dfrac{\rho_\mathrm{s}}{\dfrac{r}{r_\mathrm{s}}\left(1+\dfrac{r}{r_\mathrm{s}}\right)^2}\,,
\end{equation}
where $\rho_\mathrm{s}=4.2\times 10^6$ $\mathrm{M}_\odot/\mathrm{kpc}^3$, and $r_\mathrm{s}=24.54$ kpc. The halo virial radius and mass, defined with overdensity $\Delta_\mathrm{vir}=99.2$ with respect to the critical density, are $r_\mathrm{vir}=280.6$ kpc and $M_\mathrm{vir}=10^{12.1} \mathrm{M}_\odot$. To guarantee a finite halo mass, we apply an exponential cutoff to the halo density profile at $r_\mathrm{vir}$. Convergence test results show that such an exponential cutoff at $r\gtrsim10r_s$ brings negligible influence to the SIDM halo gravothermal evolution. The properties of this halo are chosen to be consistent with the host halo from the CDM zoom-in simulation presented in \cite{2020ApJ...896..112N}, which has also been resimulated in velocity-dependent SIDM models.\footnote{This host halo was selected from the CDM cosmological simulation \textsc{c125–1024}, and its corresponding zoom-in simulation was first presented in \cite{2015ApJ...810...21M}.} 
However, our selection of these halo properties is an arbitrary choice for testing the performance of the gravothermal solution mapping method introduced in Section~\ref{sec:mapping}. Since the gravothermal fluid formalism is scale free (i.e. independent of $\rho_s$ and $r_s$ besides some re-scaling factors), comparisons between the gravothermal solutions and isolated SIDM halo simulations of arbitrary sizes and masses are equivalent.\par 
To avoid extremely long halo core collapsing timescales and, therefore, expensive simulations, we select a large constant cross section factor $\sigma_\mathrm{c}=30$ cm$^2$/g and consider a simple velocity-dependent cross section model introduced in \cite{2021MNRAS.507.2432G},
\begin{equation}\label{eq:Gilman2021}
    \sigma(v)=\dfrac{\sigma_c}{\left(1+\dfrac{v^2}{\omega^2}\right)^2}\,,
\end{equation}
where $\omega$ is a characteristic velocity beyond which the cross section drops as $v^{-4}$. This asymptotic behavior is motivated by Coulomb or Yukawa interactions \citep{2009PhRvD..79b3519A,2009JCAP...07..004F,2010PhLB..692...70I,2018MNRAS.474..388K}. The factor 1 in the denominator is used to avoid divergent cross section for low-velocity particles in the numerical simulations.\par
We consider four different cross section models where $\omega=\infty$ (i.e. $\sigma=\sigma_c=30$ cm$^2$/g is a constant), and $\omega=1000,\ 500,\ 400$ km/s. Notice that since DM particle 1D root-mean-square (rms) velocity at the halo center and at the core-formation moment (when the halo central density reaches minimum) is about 100 km/s, corresponding to a 3D rms relative velocity of about 250 km/s, those four scenarios corresponds to constant cross section, and weak, intermediate, strong cross section velocity dependency, respectively.\par 
We use \textsc{SpherIC} \citep{2013MNRAS.433.3539G} to generate five initial conditions with identical resolutions and halo properties, but different random seeds. Through this we can quantify the effects of discreteness noise in the initial conditions on the evolution of the halo in the N-body simulations. We set the particle mass to $10^6\ \mathrm{M}_\odot$ so the halo in each simulation is resolved with $1.7\times10^6$ particles. For each cross section scenario we simulate five halos with the \textsc{Arepo} SIDM module, and we direct the reader to \cite{2012MNRAS.423.3740V,2014MNRAS.444.3684V,2019MNRAS.484.5437V} for a detailed introduction to the numerical method. This SIDM simulation pipeline has been adopted by many previous works (e.g. \citealt{2022MNRAS.513.4845Z}) and has been proved to generate valid simulation results for cross section up to 60 cm$^2$/g.\par

\section{gravothermal fluid formalism}\label{sec:gravoIntro}
Consider a spherically symmetric and isolated halo. Assuming gravothermal equilibrium is achieved at every radius and every moment, the halo evolution can be described by the following four equations (see also \citealt{2002ApJ...568..475B,Essig2019}):
\begin{equation}\label{eq:1}
    \dfrac{\partial M}{\partial r}=4\pi r^2\rho,
\end{equation}
\begin{equation}\label{eq:2}
    \dfrac{\partial(\rho v_\mathrm{rms}^2)}{\partial r}=-\dfrac{GM\rho}{r^2},
\end{equation}
\begin{equation}\label{eq:3}
    \dfrac{L}{4\pi r^2}=-\kappa\dfrac{\partial T}{\partial r},
\end{equation}
\begin{equation}\label{eq:4}
    \dfrac{\rho v_\mathrm{rms}^2}{\gamma-1}\left(\dfrac{\partial}{\partial t}\right)_M\ln\dfrac{v_\mathrm{rms}^2}{\rho^{\gamma-1}}=-\dfrac{1}{4\pi r^2}\dfrac{\partial L}{\partial r}.
\end{equation}
Throughout this paper we use $M(r)$ to denote halo enclosed mass at radius less than $r$. Then, $\rho(r)$, $v_\mathrm{rms}(r)$, $L(r)$, $\kappa(r)$, and $T(r)=mv_\mathrm{rms}^2(r)/k_\mathrm{B}$ correspond to halo density, 1D rms velocity averaged over the Maxwell-Boltzmann (MB) distribution, luminosity, thermal conductivity, and temperature at radius $r$ respectively. Here, $m$ is the DM particle mass and $k_\mathrm{B}$ is the Boltzmann constant. Since we do not consider the halo center movement throughout this work, the rms velocity and velocity dispersion among DM particles are identical at all radii. Assuming the DM particles act as a monotonic ideal gas, we set the adiabatic index $\gamma=5/3$. The Jeans equation (Eq~\ref{eq:2}) forces gravothermal equilibrium where the outward pressure $\partial(\rho v_\mathrm{rms}^2)/\partial r$ is balanced by the gravitational force. Eq~\ref{eq:3} defines the heat flux, and Eq~\ref{eq:4} is the first law of thermal dynamics. Together Eqs~\ref{eq:3}--\ref{eq:4} govern the quasi-static time evolution of a SIDM halo.\par
For ideal gas environments where the heat conduction is solely realized by the thermal motion and collisions among gas particles, the kinetic theory of gases gives heat conductivity:
\begin{equation}\label{eq:kappa_smfp}
    \kappa_\mathrm{smfp}\propto \dfrac{n\lambda^2k_\mathrm{B}}{t_r}\,,
\end{equation}
and it is conventional to define $\kappa_\mathrm{smfp}=2.1v_\mathrm{rms}k_\mathrm{B}/\sigma$, derived from integration over the MB velocity distribution. Here, $n=\rho/m$ is the local DM particle number density, $\lambda=1/(n\sigma)$ is the mean free path, and $t_r=\lambda/v_{12}$ is the relaxation time. Here we use $\sigma$ to denote DM particle scattering cross section, which is either a constant or a velocity dependent quantity. We use $v_{12}$ to distinguish particle relative velocity from the 1D-rms velocity $v_\mathrm{rms}$ used for the temperature definition.\par
Eq~\ref{eq:kappa_smfp} does not apply to low-density and low self-interaction-cross-section gravothermal systems, where most of the particles tend to orbit several times before colliding with one another. Using the Jeans' length $H=\sqrt{v_\mathrm{rms}^2/(4\pi G\rho)}$ to approximate particles' characteristic orbital radii, Eq \ref{eq:kappa_smfp} is only accurate when $\lambda\ll H$ (also referred as the short-mean-free-path or smfp regime). In this work we only care about halo evolution processes that satisfy $\lambda\gg H$, also referred as the long-mean-free-path or lmfp regime. To derive the lmfp regime heat conductivity one just needs to replace the mean free path $\lambda$ with the scale height $H$ \citep{1987degc.book.....S}:
\begin{equation}
    \kappa_\mathrm{lmfp}\propto \dfrac{nH^2k_\mathrm{B}}{t_r}\,.
\end{equation}
It is conventional to assume $\kappa_\mathrm{lmfp}=(3\beta/2)nH^2k_\mathrm{B}/t_r=0.27\beta nv_{rms}^3\sigma k_\mathrm{B}/(\mathrm{G}m)$, where $\beta$ is an adjustable parameter to match the fluid numerical solutions with N-body simulations. The value of $\beta$ cannot be derived from first-principles and is therefore treated as a free parameter to be calibrated against N-body simulations. It is known that calibration to N-Body simulations results in $\beta$ of order unity, but with some variation in the range $0.5-1.5$ depending on the specific simulation and cross-section parameters \citep{Essig2019}. We will show in Section \ref{sec:matchArepo} that in the lmfp regime, it is valid to absorb constant factors resulting from integration over the MB velocity distributions into $\beta$ when $\sigma$ is constant. However, we will demonstrate that a more careful treatment is required for relative velocity dependent cross section $\sigma(v_{12})$ cases.\par  
A nice feature of this gravothermal fluid formalism is that, given the halo initial density profile as NFW with characteristic radius $r_\mathrm{s}$ and density $\rho_\mathrm{s}$, the equation set Eq~\ref{eq:1}-\ref{eq:4} can be rewritten into the scale-free form such that the unit-less gravothermal solutions can be applied to an arbitrary halo by re-scaling. Specifically, all physical quantities in the fluid formalism equation set $x$ can be separated into a unit factor $x_0$ and a scale-free factor $\hat{x}$ (i.e. $x=x_0\hat{x}$) such that Eq~\ref{eq:1}--\ref{eq:4} can be written into the unit-less form:
\begin{equation}\label{eq:eqset}
    \begin{split}
        \dfrac{\partial \hat{M}}{\partial \hat{r}}&=\hat{\rho}\hat{r}^2\,,\\
        \dfrac{\partial(\hat{\rho}\hat{v}_\mathrm{rms}^2)}{\partial \hat{r}}&=-\dfrac{\hat{M}\hat{\rho}}{\hat{r}^2}\,,\\
        \dfrac{\hat{L}}{\hat{r}^2}&=-\hat{\kappa}\dfrac{\partial\hat{v}_\mathrm{rms}^2}{\partial \hat{r}}\,,\\
        \hat{\rho}\hat{v}_\mathrm{rms}^2\left(\dfrac{\partial}{\partial\hat{t}}\right)_{\hat{M}}\ln\dfrac{\hat{v}_\mathrm{rms}^3}{\hat{\rho}}&=-\dfrac{1}{\hat{r}^2}\dfrac{\partial\hat{L}}{\partial\hat{r}}\,.
    \end{split}
\end{equation}
The unit variables are \citep{Essig2019}:
\begin{equation}
    \begin{split}
        &r_0=r_s\,, \rho_0=\rho_s\,, M_0=4\pi r_s^3\rho^s\,,\\ 
        &v_{\mathrm{rms},0}=\sqrt{4\pi G\rho_s}r_s\,, t_0=1/\sqrt{4\pi G\rho_s}\,,\\ &L_0=(4\pi)^{5/2}G^{3/2}\rho_s^{5/2}r_s^5\,, (\sigma/m)_0=1/(\rho_s r_s)\,,
    \end{split}
\end{equation}
and it can be derived that $\hat{\kappa}_\mathrm{lmfp}=0.27\times4\pi\beta\hat{\rho}\hat{v}_\mathrm{rms}^3(\hat{\sigma/m})$, $\hat{\kappa}_\mathrm{smfp}=2.1\hat{v}_\mathrm{rms}/(\hat{\sigma/m})$. Hereafter we will abbreviate $(\hat{\sigma/m})$ as $\hat{\sigma}$ for simplicity. We assume $\hat{\kappa}=1/(\hat{\kappa}_\mathrm{lmfp}^{-1}+\hat{\kappa}_\mathrm{smfp}^{-1})$ to join the lmfp and smfp regimes smoothly.\par
The steps to solve Eq \ref{eq:eqset} numerically are \citep{Essig2019}:
\begin{enumerate}
    \item Determine input parameters $\beta$ and $\hat{\sigma}$.
    \item Grid the halo with NFW density profile $\hat{\rho}=1/\hat{r}/(1+\hat{r})^2$ into 150 log radial bins spread uniformly in range $-2\leq\log\hat{r}\leq3$\footnote{In this work $\log$ denotes a base-10 logarithm and $\ln$ denotes a natural logarithm.}\footnote{We have tested that setting the halo outer bound at $\log\hat{r}=1,2,3$ or the inner bound at $\log\hat{r}=-3,-2$ has negligible influence on the fluid solutions.}. We assume boundary conditions $\hat{M}=\hat{L}=0$ at the inner radius and $\hat{L}=0$ at the outer radius to solve for the radial profile of $\hat{v}_\mathrm{rms}$ through the Jeans equation. Following \cite{2015ApJ...804..131P} and \cite{Essig2019}, we take values of $\hat{M}_i$ and $\hat{L}_i$ of the $i$th shell at the log radial grid $\hat{r}_i$, while $\hat{\rho}_i$, $\hat{v}_{\mathrm{rms},i}$ are taken as the averaged values between the $i-1$ and the $i^\mathrm{th}$ shell.
    \item Let the fluid code take a small time step $\mathrm{d}\hat{t}$ such that the specific energy $\hat{u}=3\hat{v}_\mathrm{rms}^2/2$ changes by a factor of no more than $10^{-3}$ among all the shells:
    \begin{equation}\label{eq:dudt}
    \dfrac{\mathrm{d}\hat{u}_i}{\mathrm{d}\hat{t}}=-\dfrac{\partial\hat{L}_i}{\partial\hat{M}_i}\,.
    \end{equation}
    \item After taking one time step forward, perturb $\hat{r}$, $\hat{\rho}$, and $\hat{v}_\mathrm{rms}$ of each shell for 10 times such that the gravothermal equilibrium (Eq \ref{eq:2}) is established again throughout the halo. During the perturbation, the mass, specific energy, and entropy $s=\ln(\hat{v}_\mathrm{rms}^3/\hat{\rho})$ of each shell are conserved.
    \item Repeat step 2.\,, 3.\ and 4.\ iteratively until core collapse.
\end{enumerate}\par
We checked that the basic assumptions made in solving Eq~\ref{eq:eqset}, i.e. (1) The halo is spherically symmetric, (2) The halo mass and scale do not vary with time, are satisfied in the idealized N-Body simulations for isolated halos. However, those assumptions may be not valid for cosmological simulations due to halo interactions, mergers and the mass accretion. \par
\begin{figure}
    \hspace{-4mm}
    \includegraphics[width=0.475\textwidth]{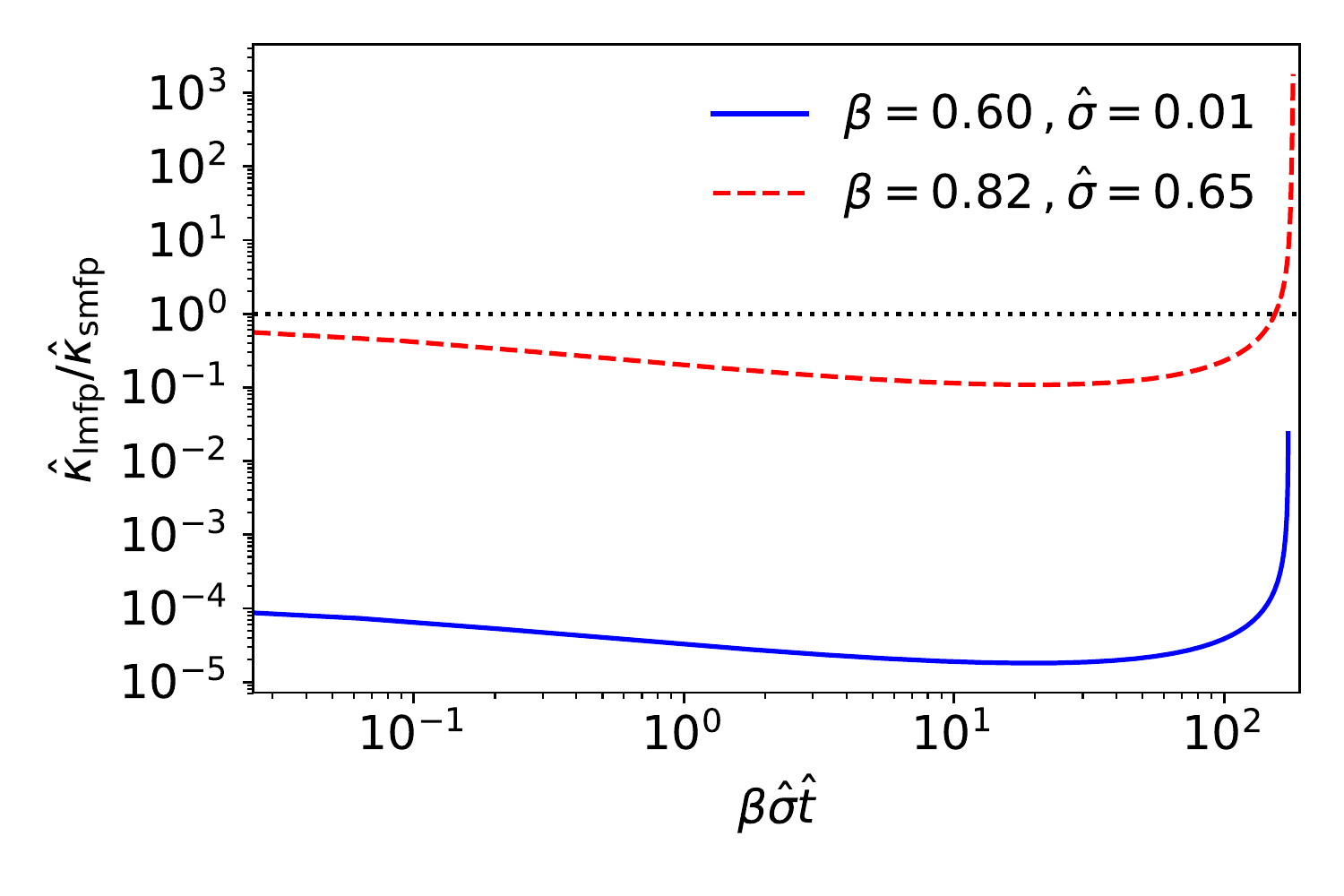}
    \caption{Time evolution of the ratio between the lmfp and the smfp heat conductivities at the halo center, assuming two different sets of parameters $\{\hat{\sigma}=0.01, \beta=0.6\}$ (blue solid) and $\{\hat{\sigma}=0.65, \beta=0.82\}$ (red dashed). For the $\{\hat{\sigma}=0.01, \beta=0.6\}$ set of gravothermal solution $\hat{\kappa}_\mathrm{lmfp}\ll\hat{\kappa}_\mathrm{smfp}$ throughout the halo evolution process, so the total heat conductivity is always dominated by the lmfp term. In the $\{\hat{\sigma}=0.65, \beta=0.82\}$ case the halo also evolves mostly in the lmfp regime, but enters the smfp regime at $\beta\hat{\sigma}\hat{t}=155$.}
    \label{fig:kappa}
\end{figure}

  
Notice that most of the current gravothermal fluid studies are interested in SIDM models with moderate cross section and near unity $\beta$ values. Such a setup results in the halo remaining in the lmfp regime for most of its life until core collapse occurs. Moreover, $\sigma$ decreases drastically with $v_\mathrm{rms}$ at $v_\mathrm{rms}\gtrsim\omega$ under the velocity-dependent cross section scenario, leaving the heat conduction more likely to be dominated by the lmfp term. We therefore focus on the gravothermal solutions in the lmfp regime in this work. Specifically, we choose input parameters $\hat{\sigma}=0.01, \beta=0.6$ in step 1. and solve Eq~\ref{eq:eqset} numerically to derive the gravothermal solutions, which will be used by the mapping method introduced in Section~\ref{sec:mappingEssig}-\ref{sec:matchArepo}. We stop the fluid code evolution when $\beta\hat{\sigma}\hat{t}$ reaches 173. At the last time step, the halo is still in the lmfp regime, and has entered the core-collapsing stage where its central density increases rapidly. The time evolution of the ratio between $\hat{\kappa}_\mathrm{lmfp}$ and $\hat{\kappa}_\mathrm{smfp}$ at the halo center is shown in Figure~\ref{fig:kappa} by the blue solid curve.\par 
We notice that the halos simulated in Section~\ref{sec:Arepo} adopt a much greater cross section of $\hat{\sigma}=0.65$. To check the validity of the lmfp assumption we numerically solve Eq~\ref{eq:eqset} for another set of input parameters $\{\beta=0.82, \hat{\sigma}=0.65\}$, and present the halo central density time evolution in the red dashed curve of Figure~\ref{fig:kappa}. This halo still evolves in the lmfp regime during most of its lifetime before core collapse, although the increase of cross section boosts the halo to be closer to the smfp regime since $\hat{\kappa}_\mathrm{lmfp}/\hat{\kappa}_\mathrm{smfp}\propto\beta\hat{\sigma}^2$. Ideally we would like to simulate a halo with smaller cross section to better match the assumption made in this work. However, this will delay the onset of core collapse and lead to more expensive and unstable idealized N-Body simulations. We will show later that even for this large constant cross section $\hat{\sigma}=0.65$ the lmfp assumption still holds, and the simulated halo evolution matches the lmfp gravothermal solution well. The smfp tension will be further alleviated when we consider the velocity-dependent cross-section scenarios, where the velocity dependency will largely decrease the effective cross section and bring the halo further away from the smfp regime.\par
\begin{figure}
    \includegraphics[width=0.45\textwidth,trim={0 0cm 0 0cm},clip]{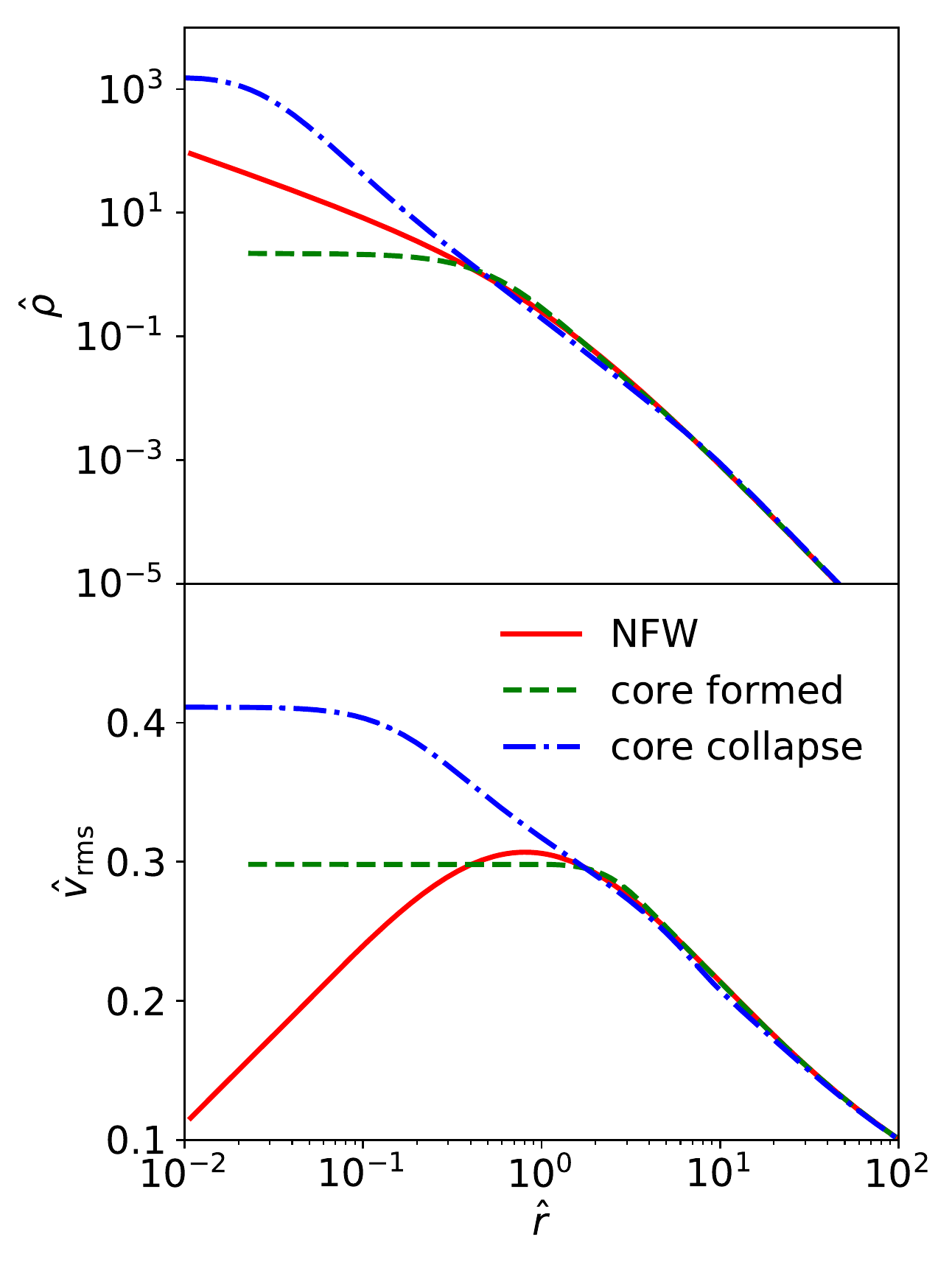}
    \caption{Gravothermal evolution phases of a SIDM halo described by the fluid formalism. The top/bottom panel shows the halo density/rms velocity profiles at different evolution stages. The red solid curves show the NFW initial condition at $\hat{t}=0$. The green dashed curves present the moment when the halo central density reaches minimum, and an isothermal core is formed at the halo center. The blue dashed-dotted curves show the afterward core collapsing process. }
    \label{fig:gravoSolEvo}
\end{figure}

SIDM halo gravothermal evolution described by Eq~\ref{eq:eqset} is shown in Figure~\ref{fig:gravoSolEvo}. Specifically, at the NFW initial state (shown in the red solid curves) where $\hat{\rho}=1/\hat{r}/(1+\hat{r})^2$, the halo temperature is highest at $\hat{r}\approx 0.8$. The heat therefore flows to both the halo center and outskirts. The halo center keeps absorbing heat until local thermal equilibrium is reached. At this point a cored region is formed where the density and the rms velocity are uniform, as shown by the green dashed curves. DM particles in the halo cored region continuously lose energy due to outward heat transfer. In this process their orbital scale decreases and the kinetic temperature increases. As shown by the blue dashed-dotted curves, the halo core density and temperature will keep increasing and will move further away from thermal equilibrium with the halo outskirts. Such a regime is also referred to as gravothermal catastrophe or core collapse.\par
While the gravothermal model is computationally much more efficient than running an N-body simulation it is still too slow to apply to large ensembles of halos as might be required in a semi-analytic model of subhalo evolution, or when generating realizations of subhalo populations for analysis of gravitational lensing systems (e.g. \cite{2012NewA...17..175B,2021MNRAS.507.2432G}). For example, the calculations shown in this section required around 24 CPU hours to compute. While this is trivially manageable for a single halo, it is impractical to apply to ensembles of millions of halos. As such it is useful to be able to map a single solution of the gravothermal model to more general cases. We will explore this possibility in the remainder of this paper.\par
Additionally, in Appendix~\ref{apx:empiricalModel} we provide a triple power law empirical model calibrated to this set of fully lmfp solutions with high precision. The advantage of this analytical model is that it captures the halo core size, core collapsing time, and the radius where the cored density profile joins smoothly to the halo NFW outskirt with simple formulas, and can be easily implemented into semi-analytic galaxy formation models.\par

\section{Mapping gravothermal solutions between Constant Cross Sections}\label{sec:mappingEssig}
In this section we briefly review the mapping method discussed in \cite{2002ApJ...568..475B,2011MNRAS.415.1125K,Essig2019} that transfers the gravothermal solutions under a constant cross section $\hat{\sigma}_1$ to another constant cross section $\hat{\sigma}_2$, assuming the halo is mostly evolving in the lmfp regime.\par
\begin{figure}
    \includegraphics[width=0.45\textwidth,trim={0 0cm 0 0cm},clip]{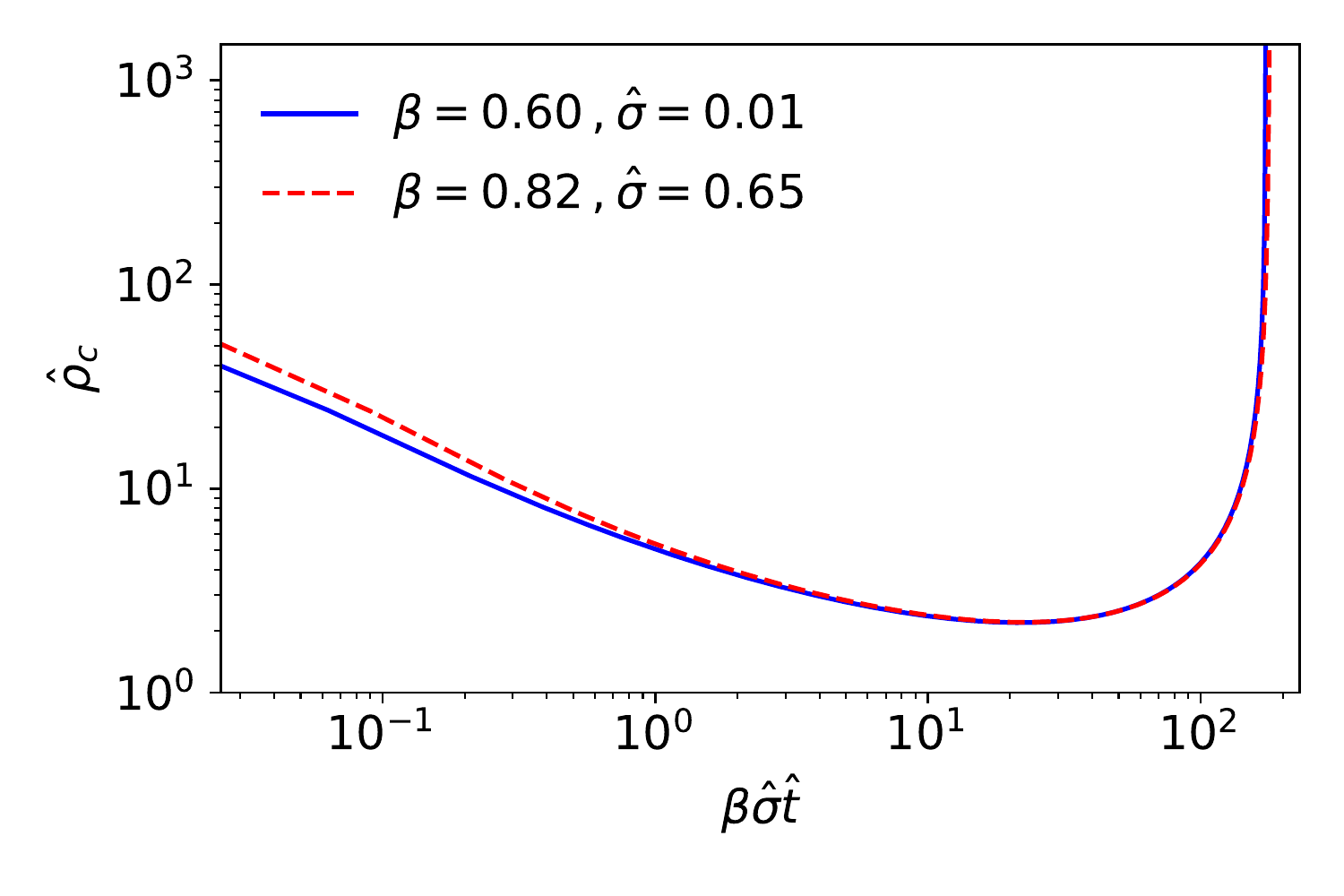}
    \caption{SIDM halo central density time evolution given by the gravothermal fluid formalism at parameters $\{\beta=0.6,\hat{\sigma}=0.01\}$ (blue solid) and $\{\beta=0.82,\hat{\sigma}=0.65\}$ (red dashed). The halo evolution is almost identical with time parameter $\beta\hat{\sigma}\hat{t}$ when the  $\kappa$ is dominated by the lmfp term.}
    \label{fig:rhoct_lmfp}
\end{figure}

The third and fourth equations in Eq~\ref{eq:eqset} shows that the halo evolution rate is degenerated with radius independent factors in the heat conductivity $\hat{\kappa}$. In the lmfp regime $\hat{\kappa}_\mathrm{lmfp}\propto \beta\hat{\sigma}$, therefore $\beta\hat{\sigma}$ and the inverse of the halo evolution time $\hat{t}$ are degenerate such that the gravothermal solution time evolution can be parameterized by $\beta\hat{\sigma}\hat{t}$. In other words, there exists a one-to-one mapping $\beta_\mathrm{a}\hat{\sigma}_\mathrm{a}\hat{t}_\mathrm{a}\leftrightarrow\beta_\mathrm{b}\hat{\sigma}_\mathrm{b}\hat{t}_\mathrm{b}$ between two sets of gravothermal solutions with parameters $\{\beta_\mathrm{a},\hat{\sigma}_\mathrm{a}\}$ and $\{\beta_\mathrm{b},\hat{\sigma}_\mathrm{b}\}$ such that: 
\begin{equation}
    \hat{x}_\mathrm{a}(\hat{r},\hat{t}_\mathrm{a})=\hat{x}_\mathrm{b}\left(\hat{r},\hat{t}_\mathrm{a}\times\dfrac{\beta_\mathrm{a}\hat{\sigma}_\mathrm{a}}{\beta_\mathrm{b}\hat{\sigma}_\mathrm{b}}\right)\,,
\end{equation}
here $\hat{x}(\hat{r},\hat{t})$ stands for arbitrary scale-free halo property at radius $\hat{r}$ and time $\hat{t}$, including $\hat{\rho}$, $\hat{v}_\mathrm{rms}$, $\hat{M}$, $\hat{L}$, etc. We show the scale-free halo central density $\hat{\rho_c}$ versus $\beta\hat{\sigma}\hat{t}$ for two sets of gravothermal solutions with very different input parameters in Figure~\ref{fig:rhoct_lmfp}. The time evolution of these two halos is almost identical after the first few time steps, where $\hat{\kappa}_\mathrm{lmfp}$ is comparable to $\hat{\kappa}_\mathrm{smfp}$ for the $\hat{\sigma}=0.65$ case, and the lmfp assumption is not very accurate (see Figure~\ref{fig:kappa}).\par

\begin{figure}
    \includegraphics[width=0.5\textwidth]{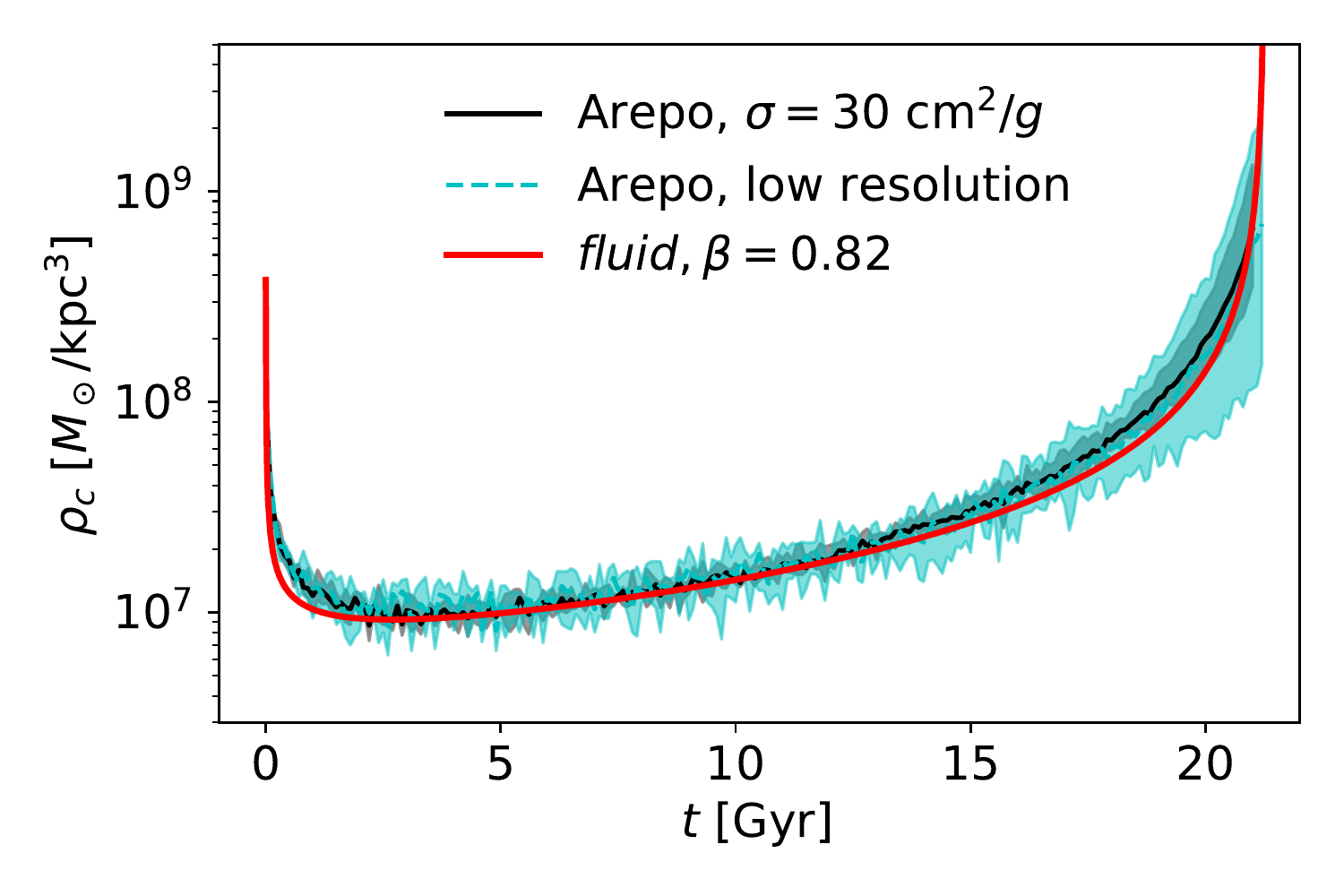}
    \caption{Halo central density evolution comparisons between the averaged Arepo N-Body simulation results (black solid curve) and the gravothermal fluid formalism numerical solutions (red solid curve). The gray band encloses the maximum and minimum halo central density among the five \textsc{Arepo} realizations. The cyan band shows convergence test results where the particle number in each N-Body realization is decreased from $1.7\times10^6$ to $4.2\times10^5$.}
    \label{fig:c30}
\end{figure}

Although the gravothermal fluid formalism considers a very simplified picture, it provides numerical solutions in surprisingly good agreement with idealized N-Body simulations of isolated halos up to a slightly varying $\beta$ factor. As an example, in Figure~\ref{fig:c30} we present halo central density versus evolution time relation comparisons between the gravothermal solutions (red solid curve) and the averaged Arepo N-body simulations (black solid curve) for a Milky Way sized SIDM halo with $\rho_\mathrm{s}=4.2\times10^6 \mathrm{M}_\odot/\mathrm{kpc}^3$, $r_\mathrm{s}=24.54$ kpc, and $\sigma/m=30$ cm$^2$/g (see Section~\ref{sec:Arepo} for simulation details). To test convergence, we reduce the particle number by a factor of 4 in each N-Body realization and present the averaged simulation results as the cyan dashed curve. We find the rescaled lmfp gravothermal solution with $\beta=0.82$ captures the overall halo evolution and reproduces the core collapsing time in the Arepo simulation results. We notice that the fluid numerical solution has halo central density that drops faster than the N-Body results prior to isothermal core formation ($0\leq t/\mathrm{Gyr}\leq2$). The central density also grows slightly faster than N-Body at $t\gtrsim18$ Gyr. Those differences are, again, caused by the fact that we have selected a very large cross section for the N-Body simulations. The basic assumption of this mapping method, that the halo is always in the lmfp regime, is not of high accuracy. Since our focus is to extend the lmfp gravothermal fluid formalism to velocity-dependent cross section scenario and capture the overall halo evolution in the N-Body simulations, in this work we will accept this slight disagreement. Moreover, the smfp problem is alleviated when we simulate velocity dependent cross section models.\par

\section{Mapping gravothermal solutions from constant to velocity-dependent cross section}\label{sec:mapping}
In this section we introduce a mapping method to transfer the gravothermal solutions under a constant cross section $\hat{\sigma}$ to those under a velocity-dependent cross section model $\sigma(v)$. In numerical simulations the velocity used to determine the scattering probability between two particles is generally the particle relative velocity $v_{12}$, but in this section we will consider a simpler case where the cross section is determined by the particle rms velocity. We will discuss this mapping method under the more practical $\sigma(v_{12})$ scenario in Section \ref{sec:matchArepo}.\par 
In this paper we will use the cross section model introduced by Eq~\ref{eq:Gilman2021} as an example, but this mapping method is applicable to arbitrary $\sigma(v)$ model.\par
In Section~\ref{sec:mappingEssig} we have shown that there exists a one-to-one mapping $\beta_\mathrm{a}\hat{\sigma}_\mathrm{a}\hat{t}_\mathrm{a}\leftrightarrow\beta_\mathrm{b}\hat{\sigma}_\mathrm{b}\hat{t}_\mathrm{b}$ between two sets of gravothermal solutions with different $\beta$ and constant cross sections. Now consider two sets of gravothermal solutions, one with constant cross section $\{\beta_\mathrm{a},\hat{\sigma}_\mathrm{a}\}$ and the other with a velocity-dependent cross section $\{\beta_\mathrm{b},\hat{\sigma}_\mathrm{b}(\hat{v}_{\mathrm{rms, b}})\}$. The linear time mapping does not hold anymore since $\hat{\sigma}_\mathrm{b}(\hat{v}_{\mathrm{rms, b}})$ is a function of radius and time rather than a constant. However, if in every small time interval we may assume a characteristic halo-wide cross section (or, equivalently,  a characteristic velocity) such that we can drop  the cross section radial dependence $\hat{\sigma}(\hat{v}_{\mathrm{rms, b}}(\hat{r},\hat{t}_\mathrm{b}))\approx \hat{\sigma}(\hat{v}_{\mathrm{rms, b}}^c(\hat{t}_\mathrm{b}))$, then there exists the following one-to-one mapping for every small time step:
\begin{equation}\label{eq:tMap}
\begin{split}
    \beta_\mathrm{a}\hat{\sigma}_\mathrm{a}d\hat{t}_a&=\beta_\mathrm{b}\hat{\sigma}_\mathrm{b}(\hat{v}_{\mathrm{rms, b}}^c(\hat{t}_\mathrm{b}))\mathrm{d}\hat{t}_\mathrm{b}\\
    &=\beta_\mathrm{b}\hat{\sigma}_\mathrm{b}(\hat{v}_{\mathrm{rms, a}}^c(\hat{t}_\mathrm{a}))\mathrm{d}\hat{t}_\mathrm{b}\,.
\end{split}
\end{equation}
Here the second equation holds because alternating a time dependent while radius independent SIDM cross section has no influence on the gravothermal fluid solution besides non-linearly stretching the time axis. In other words there still exists the following one-to-one mapping between these two sets of gravothermal solutions:
\begin{equation}
\begin{split}
    \hat{x}_\mathrm{a}(\hat{r},\hat{t}_\mathrm{a})&= \hat{x}_\mathrm{b}(\hat{r},\hat{t}_\mathrm{b})\\
    &=\hat{x}_\mathrm{b}\left(\hat{r},\int_0^{\hat{t}_\mathrm{a}}\dfrac{\beta_\mathrm{a}\hat{\sigma}_\mathrm{a}}{\beta_\mathrm{b}\hat{\sigma}_\mathrm{b}(\hat{v}_{\mathrm{rms, a}}^c(\hat{t}'_\mathrm{a}))}\mathrm{d}\hat{t}'_\mathrm{a}\right)\,.
\end{split}
\end{equation}
In practice one has the constant cross section gravothermal solution that specifies the halo density and rms velocity radial profiles at an array of time $\hat{t}_{\mathrm{a}}=\{\hat{t}^{0}_{\mathrm{a}},\hat{t}^{1}_{\mathrm{a}},...,\hat{t}^{N}_{\mathrm{a}}\}$ (here $\hat{t}^{i}_{\mathrm{a}}$ is the time of the $i^{\mathrm{th}}$ step), and aims to map it to some velocity-dependent cross section scenario. To achieve this one will need to first compute the time step $\delta\hat{t}_{\mathrm{a}}=\{\hat{t}^{0}_\mathrm{a},\hat{t}^{1}_{\mathrm{a}}-\hat{t}^{0}_{\mathrm{a}},...,\hat{t}^{N}_{\mathrm{a}}-\hat{t}^{N-1}_{\mathrm{a}}\}$, and then use Eq~\ref{eq:tMap} to compute the mapped time step $\delta\hat{t}_{\mathrm{b}}=[\beta_\mathrm{a}\hat{\sigma}_{\mathrm{a}} / \beta_{\mathrm{b}}\hat{\sigma}_{\mathrm{b}}(\hat{v}_{\mathrm{rms, a}}^{c}(\hat{t}_{\mathrm{a}}))]\delta\hat{t}_{\mathrm{a}}$. Summing $\delta\hat{t}_{\mathrm{b}}$ up, a new time axis $\hat{t}_{\mathrm{b}}$ is derived such that $\hat{x}_{\mathrm{b}}(\hat{r},\hat{t}_{\mathrm{b}})=\hat{x}_{\mathrm{a}}(\hat{r},\hat{t}_{\mathrm{a}})$.\par 
The remaining problem is to determine how to choose the characteristic rms velocity $\hat{v}_{\mathrm{rms}}^c$ at each time. Since the SIDM core evolves the most with time, while the halo outskirt shows very little time evolution (see Figure~\ref{fig:gravoSolEvo}), we find that simply using the halo central rms velocity shows very good mapping performance:
\begin{equation}\label{eq:vch}
    \hat{v}_{\mathrm{rms}}^c(\hat{t})=\hat{v}_{\mathrm{rms}}(\hat{r}=\hat{r}_\mathrm{min},\hat{t})\,.
\end{equation}\par
Here $\hat{r}_\mathrm{min}=10^{-2}$ is the inner most radial bin of the fluid formalism.\par
\begin{figure}
    \includegraphics[width=0.5\textwidth]{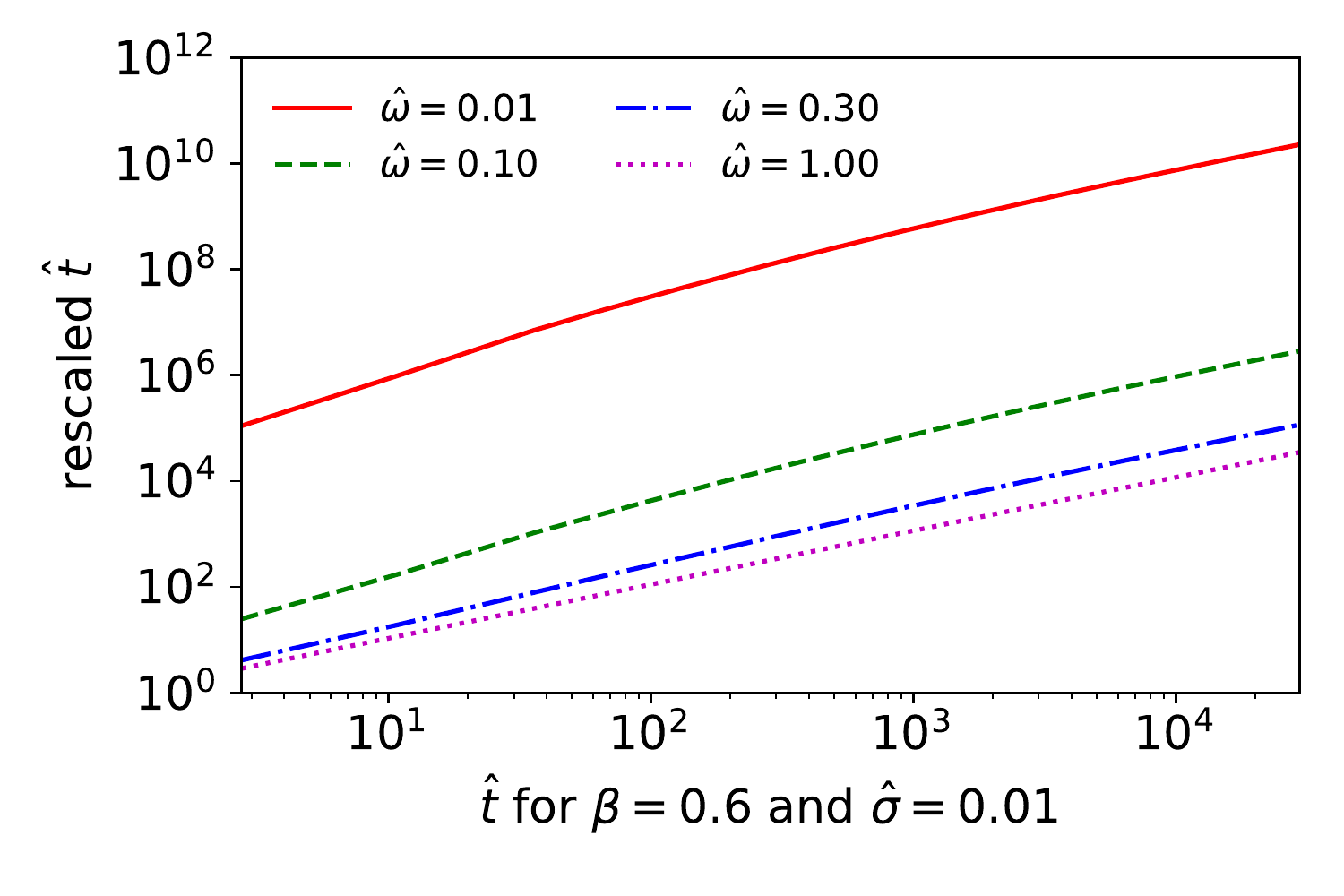}
    \caption{Mapping relation for the gravothermal solution time variable. The x axis present the scale-free time $\hat{t}$ of the gravothermal solution under constant cross section $\hat{\sigma}=0.01$. We map $\hat{t}$ to velocity-dependent cross section model $\hat{\sigma}=[\hat{\sigma}_c/(1+\hat{v}_\mathrm{rms}^2/\hat{\omega}^2)^2]$, with $\hat{\sigma}_c=0.01$ and $\hat{\omega}=0.01,0.1,0.3,1.0$. The re-scaled time is presented on the y axis for each case. For simplicity we have fixed $\beta=0.6$ in this test.}
    \label{fig:tt}
\end{figure}

 
To test this mapping method, we assume $\hat{\sigma}_\mathrm{c}=0.01$, $\beta=0.6$, and numerically solve Eq~\ref{eq:eqset} for two cross section models. For the first test we assume a constant cross section $\hat{\sigma}=\hat{\sigma}_\mathrm{c}$, while for the second test we assume $\hat{\sigma}=[\hat{\sigma}_c / (1+\hat{v}_\mathrm{rms}^2/\hat{\omega}^2 )^2 ]$. We consider four velocity-dependent cross section models with $\hat{\omega}=0.01,0.1,0.3,1.0$, corresponding to very strong, strong, intermediate, and week velocity dependency. Figure~\ref{fig:tt} presents the unitless time of the constant cross section gravothermal solution re-scaled by this mapping method. Notice that the dark matter particle 1D velocity dispersion $\hat{v}_\mathrm{rms}$ varies from 0.1 to 0.4 near the halo center (see Figure~\ref{fig:gravoSolEvo} bottom panel), which is small comparing to 1.0. The halo SIDM cross section for the $\hat{\omega}=1.0$ case is therefore almost a constant $\hat{\sigma}=\hat{\sigma}_c$, corresponding to a nearly trivial time mapping relation shown by the magenta dotted curve. Decreasing $\hat{\omega}$ suppresses the total cross section value $\hat{\sigma}$, and also boosts the velocity dependence of the particle scattering probabilities. As a result, the re-scaled $\hat{t}$ enlarges for the lower $\hat{\omega}$ cases, and the time mapping non-linearity becomes more significant.\par
\par

\begin{figure*}
    \centering
    \includegraphics[width=0.8\textwidth]{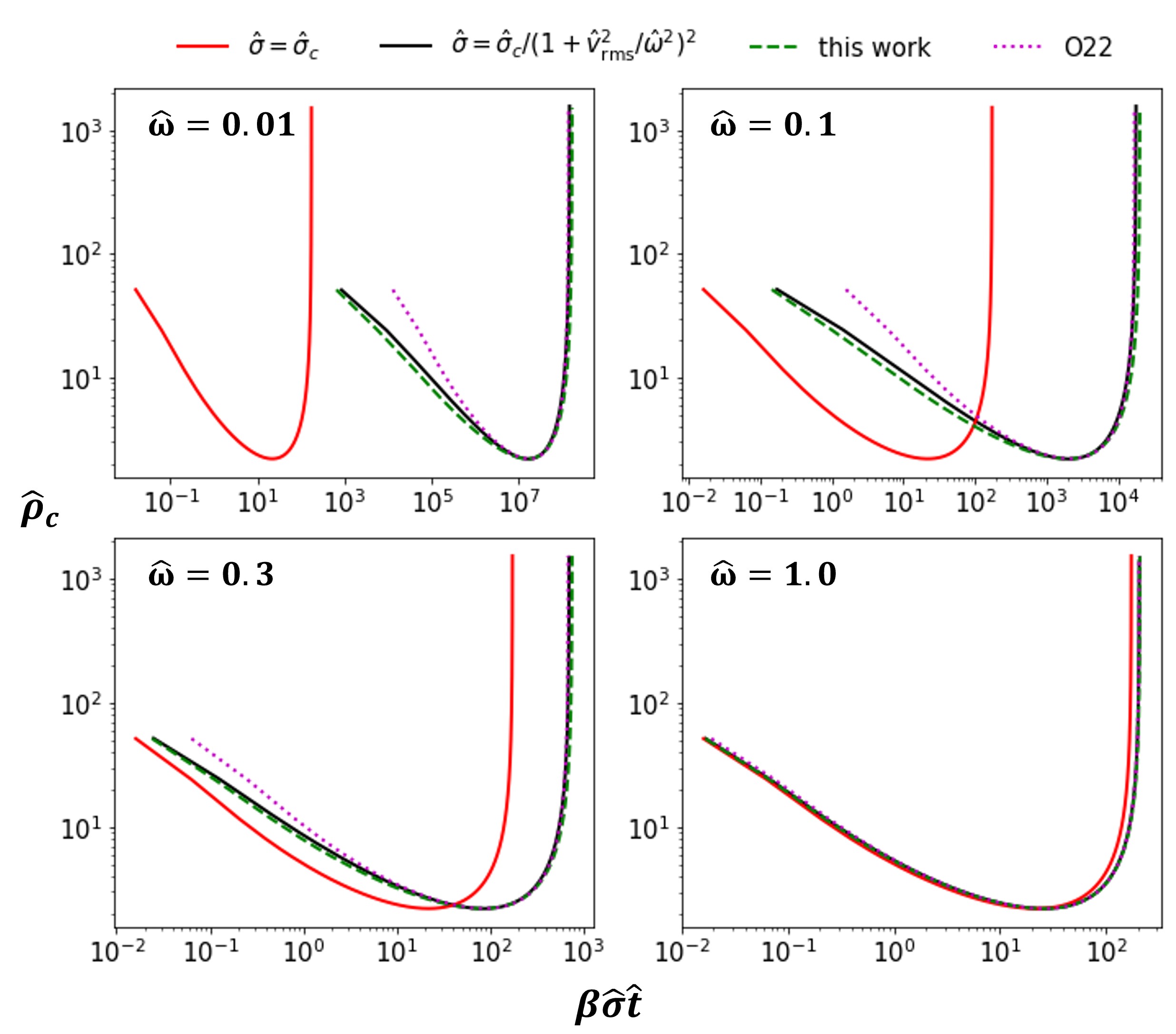}
    \caption{Scale-free SIDM halo central density evolution under a constant cross section (red solid) and a velocity-dependent cross section model (black solid). In each panel the re-scaled constant cross section gravothermal solutions given by the method introduced in Section~\ref{sec:mapping} and \citetalias{2022arXiv220406568O} are presented in the green dashed curve and magenta dotted curve, respectively. We consider four different cross section models with $\hat{\omega}=0.01,0.1,0.3,1.0$ and compare the numerical solution with the re-scaled results in the top left, top right, bottom left, and bottom right panels. Both mapping methods show reasonable agreement with the gravothermal numerical solutions. We notice that \citetalias{2022arXiv220406568O} is slightly more accurate in mapping the core collapsing time, while this work more accurately reproduces the halo evolution before its central density reaches a minimum.}
    \label{fig:gravoMap}
\end{figure*}

Comparisons between the mapped constant cross section gravothermal solutions (green dashed lines) and the actual velocity dependent gravothermal solutions (black solid curves) are shown in Figure~\ref{fig:gravoMap}. We find this simple remapping method can successfully reproduce the gravothermal solutions under velocity-dependent cross section models in all the cases, although it over-estimates the core collapsing time by 14\%, 13\%, 7.5\%, and 1.3\%, respectively. This overestimation is caused by Eq~\ref{eq:vch} where we set the halo central rms velocity as a characteristic velocity that determines the overall DM particle cross section at every moment. A true characteristic $v_\mathrm{rms}$ is generally lower than the central velocity, and therefore our method slightly underestimates the effective cross section at each time and will further overestimate the core collapsing time.\par 
We also check the performance of the mapping method introduced in \citetalias{2022arXiv220406568O}. Instead of using a time-varying characteristic rms velocity $\hat{v}_\mathrm{rms}^\mathrm{c}$, \citetalias{2022arXiv220406568O} makes a more simplified assumption to use the rms velocity at the halo center and at the instance of maximum core size as the characteristic value. Effectively the characteristic rms velocity is assumed to be:
\begin{equation}\label{eq:vchO}
    \hat{v}_{\mathrm{rms}}^\mathrm{c}=\hat{v}_\mathrm{rms}(\hat{r}=\hat{r}_\mathrm{min},\hat{t}=\hat{t}_{\hat{\rho}_{c,\mathrm{min}}})\,.
\end{equation}
We show the halo central density time evolution mapped by \citetalias{2022arXiv220406568O} as magenta dotted curves in Figure~\ref{fig:gravoMap}. We show that the \citetalias{2022arXiv220406568O} method only underestimates the core collapsing time by 6\%, 5\%, 2.3\%, and 0.3\% for the $\hat{\omega}=0.01,0.1,0.3,1.0$ scenarios, but it cannot correctly capture the halo evolution prior to the instance of maximum core size.\par  
The accuracy of our mapping method can be further boosted if we compute the characteristic $\hat{v}_\mathrm{rms}^\mathrm{c}$, or, equivalently, the characteristic cross section, at each time step more carefully. Specifically, at every time step we can define the characteristic cross section as the averaged cross section weighted by the mass of multiple shells:
\begin{equation}\label{eq:avgSigma}
    \hat{\sigma}^\mathrm{c}=\dfrac{\Sigma_0^{i_\mathrm{out}}4\pi \hat{r}_i^2\delta\hat{r}_i\hat{\rho}_i\hat{\sigma}_{b,c}(\hat{v}_\mathrm{rms,i})}{\Sigma_0^{i_\mathrm{out}}4\pi\hat{r}_i^2\delta\hat{r}_i\hat{\rho}_i}\,,
    \vspace{1mm}
\end{equation}
where $i$ is the shell index, and $i_\mathrm{out}$ corresponds to the shell that joins smoothly to the halo NFW outskirts, beyond which the DM particle self-interaction is negligible. The radius of shell $i_\mathrm{out}$ is given by Eq~\ref{eq:ro_pro} and Eq~\ref{eq:ro_pos} in Appendix~\ref{apx:empiricalModel}. After this step we can map the time step of the constant cross section gravothermal solution as $\delta\hat{t}_\mathrm{b}=[ \beta_\mathrm{a}\hat{\sigma}_\mathrm{a} / \beta_\mathrm{b}\hat{\sigma}_\mathrm{b}^c(\hat{t}_\mathrm{a}) ] \delta\hat{t}_\mathrm{a}$. We find that this more careful way of computing the characteristic cross section effectively decreases the mapping method fractional error in predicting the core collapsing moment to 6.4\%, 5.6\%, 2.9\%, and 0.5\% for $\hat{\omega}=0.01,0.1,0.3,1.0$ cases, providing performance equally as good as \citetalias{2022arXiv220406568O} while maintaining a good match to the actual gravothermal numerical solutions before the instance of maximum core size. However, we do not use this more complicated treatment as the default method because it can be computationally expensive when applied to a large number of halos. Furthermore, in practice the at most $\sim10\%$ core collapsing time mapping error is small compared to the $\beta$ factor uncertainties.\par 

\section{Calibrating the mapping method to N-Body simulation}\label{sec:matchArepo}
To calibrate the gravothermal solution remapping method introduced in Section~\ref{sec:mapping} to N-body simulations, we consider cross section models that are determined by the relative velocity $v_{12}$ between two scattering particles. Such velocity-dependent models will influence the heat conductivity averaged over the MB velocity distribution $\langle\kappa_\mathrm{lmfp}\rangle$, resulting in additional~complexity.\par
It is worth recalling how $\langle\kappa_\mathrm{lmfp}\rangle$ is computed for the constant cross section case. By convention:
\begin{equation}\label{eq:kappav}
    \kappa_\mathrm{lmfp}(v_{12})=\dfrac{3}{2}\beta nH^2\dfrac{k_\mathrm{B}}{t_r}=\dfrac{3\beta k_\mathrm{B}}{8\pi \mathrm{G}}\dfrac{n\sigma}{m} v^2_\mathrm{rms}v_{12}\,,
\end{equation}
and the relative-velocity can be defined as $v_{12}=\sqrt{v_1^2+v_2^2-2v_1v_2\cos\theta_{12}}$, where $\theta_{12}$ is the angle between the velocity vectors $\boldsymbol{v}_1$ of particle 1 and $\boldsymbol{v}_2$ of particle 2. To compute the  averaged heat conductivity over the MB velocity distribution, previous works compute the averaged $v_{12}$. Specifically, let us define the MB~distribution and the averaged $v_{12}$ as:
\begin{widetext}
\begin{equation}
\begin{split}
    f_{MB}(v)\mathrm{d}^3v&=f_\mathrm{MB}(v)v^2\mathrm{d}v\sin(\theta)\mathrm{d}\theta \mathrm{d}\phi=v^2\left(\dfrac{m}{2\pi k_\mathrm{B}T}\right)^{3/2}\exp\left(-\dfrac{mv^2}{2k_\mathrm{B}T}\right)\sin(\theta)dv\mathrm{d}\theta \mathrm{d}\phi\,,\\
    \langle v_{12}\rangle&=\int_0^\pi \sin\theta_1\mathrm{d}\theta_1\int_0^{2\pi}\mathrm{d}\phi_1\int_0^{2\pi}\mathrm{d}\phi_{12}\int_0^\infty v_1^2f_\mathrm{MB}(v_1)\mathrm{d}v_1\int_0^\infty v_2^2f_\mathrm{MB}(v_2)\mathrm{d}v_2\int_0^\pi v_{12}\sin(\theta_{12})\mathrm{d}\theta_{12}\\
    &=\dfrac{4}{\sqrt{\pi}}\sqrt{\dfrac{k_\mathrm{B}T}{m}}=\dfrac{4}{\sqrt{\pi}}v_\mathrm{rms}=2.26v_\mathrm{rms}\,.
\end{split}
\end{equation}
\end{widetext}
Using this result for $\langle v_{12}\rangle$ in Eq~\ref{eq:kappav} gives the  definition assuming a constant DM particle cross section:
\begin{equation}\label{eq:kappalmfp}
    \langle\kappa_\mathrm{lmfp}\rangle=\dfrac{3\beta k_\mathrm{B}}{8\pi \mathrm{G}}\dfrac{n\sigma}{m} v^2_\mathrm{rms}\langle v_{12}\rangle=0.27\beta nv_\mathrm{rms}^3\sigma k_\mathrm{B}/(\mathrm{G}m).
\end{equation}

One may argue that a better definition of the averaged heat conductivity might be $\langle\kappa_\mathrm{lmfp}\rangle\propto\langle v_1v_2v_{12}\rangle$, or $\langle v_{12}^3\rangle$, or $\langle v^n\rangle/v_\mathrm{rms}^{n-3}$, where $n$ is an arbitrary number. The specific choice does not matter for constant cross section scenarios because those different integrations will only cause a constant factor difference to Eq~\ref{eq:kappalmfp}, which can be absorbed into the free parameter $\beta$. This is no longer true if a velocity-dependent cross section is assumed. Specifically, the definition $\langle\kappa_\mathrm{lmfp}\rangle\propto\langle v^n\sigma(v_{12})\rangle/(v_\mathrm{rms}^{n-3}\sigma_c)$ results in $\langle v^n\rangle/v_\mathrm{rms}^{n-3}$ for $\omega\rightarrow\infty$. However, with finite $\omega$ the $v^n$ term serves as a weight in the integration. Larger $n$ biases the integration over the MB distribution toward higher velocities, corresponding to a smaller effective cross section and longer halo core collapse timescale. Therefore it is necessary to define $\langle\kappa_\mathrm{lmfp}\rangle$ in a physically motivated way that matches N-Body results.\par
Notice that the heat flux is determined by the product of the conductivity and the the temperature gradient: $H\partial T / \partial r \sim\Delta T\sim\Delta E$ is the characteristic temperature variation and energy transferred within a particle's orbit. Here temperature is also a statistical property averaged over the MB distribution $T\propto\langle v_{12}^2\rangle$, so the energy transfer rate is proportional to $v_{12}^3\sigma(v_{12})$ \citep{2021PhRvD.103c5006C}. We therefore define the averaged heat conductivity as:
\begin{equation}\label{eq:kappalmfp_v12}
    \langle\kappa_\mathrm{lmfp}\rangle=0.27\beta nv_\mathrm{rms}^3 k_\mathrm{B}/(\mathrm{G}m)\dfrac{\langle v_1v_2v_{12}^3\sigma(v_{12})\rangle}{\langle v_1v_2v_{12}^3\rangle}\,.
\end{equation}
Here the $v_1v_2$ term in the numerator comes from the scale height $H^2$. One $v_{12}$ term is contributed by $1/t_\mathrm{r}$, and the other $v_{12}^2$ term comes from the temperature gradient. The numerator is defined such that Eq~\ref{eq:kappalmfp_v12} reduces to Eq~\ref{eq:kappalmfp} at $\omega\rightarrow\infty$. This definition is slightly different from the ansatz assumed in \citetalias{2022arXiv220406568O}, where $\langle\kappa_\mathrm{lmfp}\rangle\propto\langle v_{12}^3\sigma(v_{12})\rangle$. We notice that the triple integral $\langle v_1v_2v_{12}^3\sigma(v_{12})\rangle$ needs to be computed for every time interval, which significantly slows down the mapping calculations. However, it can be shown that the integration results are only sensitive to $\omega/\sqrt{T}\propto\omega/v_\mathrm{rms}$ times $\beta\sigma_c$, we therefore create a lookup table for the integration result spanning in parameter range $0.01\leq\omega/v_\mathrm{rms}\leq100$ and use the linearly interpolated integration results to achieve fast mapping.\par
To map a set of constant cross section gravothermal solutions with parameters $\{\beta_\mathrm{a},\hat{\sigma}_{c,a}\}$ to those assuming a velocity-dependent cross section model with parameters $\{\beta_\mathrm{b},\hat{\sigma}_{c,b},\hat{\omega}\}$, we perform the following steps:\par
\begin{figure*}
    \includegraphics[width=1.0\textwidth]{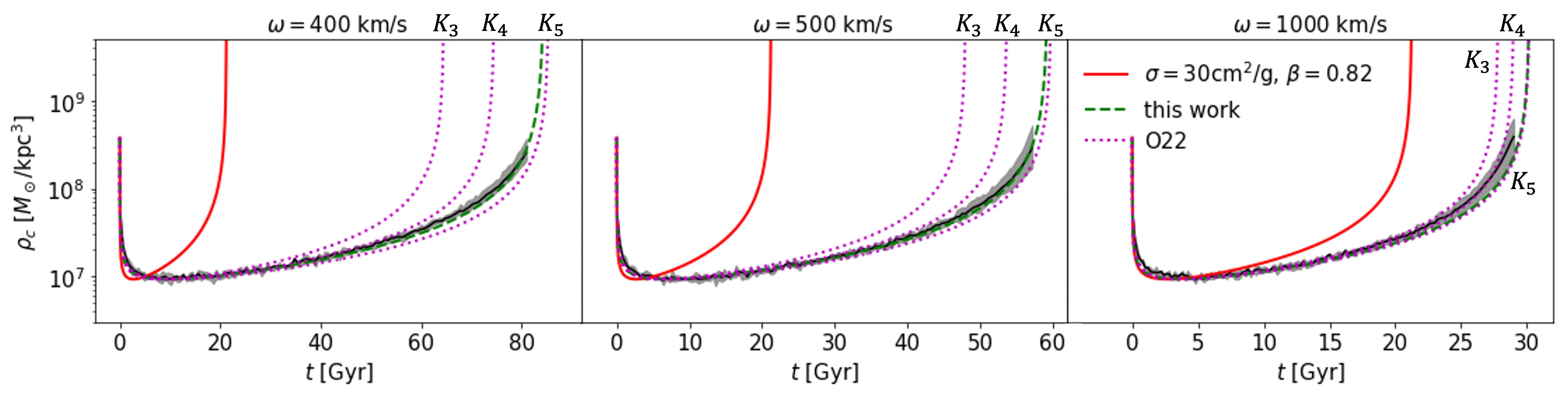}
    \caption{Comparisons of halo central density time evolution between the averaged \textsc{Arepo} N-Body simulation results (black solid curve) and the mapped gravothermal solutions (green dashed curves) under different velocity dependent cross section models. In each panel, the maximum and minimum halo central density among the five \textsc{Arepo} realizations are enclosed by the gray band. The red curve shows the gravothermal solution that matches with N-Body assuming a constant cross section $\sigma=30$ cm$^2$/g, which is identical to the red curve shown in Figure~\ref{fig:gravoSolEvo}. We use $\beta=0.82$ as calibrated in Figure~\ref{fig:c30}, and assume it does not vary with $\omega$. We find the mapping method introduced in this work matches with \textsc{Arepo} N-Body simulation results extremely well. It outperforms the $\langle\kappa_\mathrm{lmfp}\rangle\propto K_3$ definition, which is assumed as an ansatz in \citetalias{2022arXiv220406568O}. We also find under a linear time re-scaling, the definition of the averaged heat conductivity $\langle\kappa_\mathrm{lmfp}\rangle\propto K_5$ performs equally as well as Eq~\ref{eq:kappalmfp_v12}.}
    \label{fig:MapvsArepo}
\end{figure*}

\begin{enumerate}
\item{Compute the time step $\delta\hat{t}_\mathrm{a}=\{\hat{t}^0_\mathrm{a},\hat{t}^1_\mathrm{a}-\hat{t}^0_\mathrm{a},...,\hat{t}^N_\mathrm{a}-\hat{t}^{N-1}_\mathrm{a}\}$.}
\item{Compute the mapped time step $\delta\hat{t}_\mathrm{b}=[ \beta_\mathrm{a}\hat{\sigma}_\mathrm{c,a}/\beta_\mathrm{b}][\langle v_1v_2v_{12}^3\rangle_{\hat{t}_\mathrm{a}}/ \langle v_1v_2v_{12}^3\hat{\sigma}_b(v_{12})\rangle_{\hat{t}_\mathrm{a}} ] \delta\hat{t}_\mathrm{a}$. Here $\langle x\rangle_{\hat{t}_\mathrm{a}}$ implies computing the averaged $x$ over a MB distribution where the temperature is given by the constant cross section gravothermal solutions at the innermost radius and time $\hat{t}_\mathrm{a}$. We choose the halo central temperature as a characteristic value at every time step with the same argument as discussed in Section~\ref{sec:mapping}.}
\item{Sum over all $\delta\hat{t}_\mathrm{b}$ to obtain a new time axis such that $\hat{x}_\mathrm{b}(\hat{r},\hat{t}_\mathrm{b})=\hat{x}_\mathrm{a}(\hat{r},\hat{t}_\mathrm{a})$.}
\end{enumerate}
Assuming that in the idealized N-Body simulation $\beta$ does not vary with $\omega$, we compare the mapped gravothermal solutions and \textsc{Arepo} N-Body simulations under velocity-dependent cross section models introduced by Eq~\ref{eq:Gilman2021} with different $\omega$. We find excellent matches, as presented in Figure~\ref{fig:MapvsArepo}. We notice that this comparison is only performed in a limited $\omega\geq 400$ km/s range for this Milky Way sized halo. Ideally, we would extend this comparison to cross section models with stronger velocity dependency, i.e.\ with even lower characteristic velocity $\omega$. However, a halo collapses more slowly as $\omega$ decreases, which makes the N-Body simulation less stable and computationally more expensive. Furthermore, for the Milky Way sized halo we simulate in this work $\omega\leq400$ km/s is a less interesting scenario, where the halo collapsing time is much longer than the age of the universe.\par 
We also show the gravothermal solution mapped by \citetalias{2022arXiv220406568O} in the magenta dotted curves in Figure~\ref{fig:MapvsArepo}. Specifically, \citetalias{2022arXiv220406568O} suggests to define the averaged heat conductivity as:
\begin{equation}
    \langle\kappa_\mathrm{lmfp}\rangle\propto K_n=\dfrac{\langle v_{12}^n\sigma(v_{12})\rangle}{\langle v_{12}^n\rangle}\Big|_{r_\mathrm{min},t_{\rho_{c,\mathrm{min}}}}\,, n=3\,.
\end{equation}
It is emphasized in \citetalias{2022arXiv220406568O} that the assumption $\langle\kappa_\mathrm{lmfp}\rangle\propto K_3$ should be further tested by N-Body simulations. We find this assumption overestimates the effective cross section as well as the heat conductivity, leading to a faster core collapse than the N-Body results. However, we find assuming $\langle\kappa_\mathrm{lmfp}\rangle\propto K_5$ and a linear time re-scaling gives mapping performance as precise as Eq~\ref{eq:kappalmfp_v12}. This good match is also proved independently by \cite{2022JCAP...09..077Y} through N-body simulations.\par

\section{Extend the lmfp mapping method to more general cases}\label{sec:generalization}
The lmfp gravothermal solution mapping method introduced in Section~\ref{sec:mappingEssig}-\ref{sec:matchArepo} can be easily generalized to the smfp regime as well as to the intermediate regime. However, in this work we do not test the validity of the smfp and intermediate regime mapping methods introduced in this section due to difficulties in both computation and physics. Neither will we apply these mappings in the following sections. From the computational aspect, we find the gravothermal fluid formalism introduced in Section~\ref{sec:gravoIntro} runs into numerical issues when solving systems with very large cross section. Furthermore, the \textsc{Arepo} idealized SIDM simulation pipeline introduced in Section~\ref{sec:Arepo} is not tested for isolated halos with cross section greater than 60 cm$^2$/g. From the aspect of physics, a very large SIDM cross section may change the structure formation from that of CDM. The initial density profile of the large cross section SIDM halo is also uncertain and unlikely to be NFW \citep{2003JKAS...36...89A}.\par
Let us first revisit Eq~\ref{eq:eqset} in the fully smfp regime. Now since we switch focus from $\hat{\kappa}_\mathrm{lmfp}\propto\beta\hat{\sigma}$ to $\hat{\kappa}_\mathrm{smfp}\propto1/\hat{\sigma}$, it becomes clear that the time mapping relation changes from $\beta_a\hat{\sigma}_a\hat{t}_a\leftrightarrow\beta_b\hat{\sigma}_b\hat{t}_b$ to $\hat{t}_a/\hat{\sigma}_a\leftrightarrow\hat{t}_b/\hat{\sigma}_b$ for the constant SIDM cross section models. For the relative velocity dependent cross section scenarios one only needs to modify the time interval mapping as:
\begin{equation}\label{eq:smfp_map}
    \delta\hat{t}_b=\dfrac{\langle v_{12}^3\rangle_{\hat{t}_a}/\hat{\sigma}_{c,a}}{\langle v_{12}^3/\hat{\sigma}_b(v_{12})\rangle_{\hat{t}_a}}\delta\hat{t}_a\,,
\end{equation}
to map gravothermal solutions computed under constant cross sections $\hat{\sigma}_{c,a}$ to those assuming velocity dependent cross section models $\hat{\sigma}_b(v_{12})$. Here one $v_{12}$ term in the MB integration comes from the mean free path $\lambda$ and another $v_{12}^2$ term comes from the energy transfer as has been discussed in Section~\ref{sec:matchArepo}. The two key differences between the lmfp and smfp mapping methods are: (1) The smfp heat conductivity and, therefore, the gravothermal solutions are no longer dependent on the free parameter $\beta$. (2). In the lmfp regime a halo collapses faster with a larger cross section, but the situation is reversed in the smfp regime.\par
To consider the intermediate regime where the mean free path and the scale height are comparable, the time mapping relation among constant cross section models has form $\hat{\kappa}_a\hat{t}_a\leftrightarrow\hat{\kappa}_b\hat{t}_b$, where $\hat{\kappa}=1/(\hat{\kappa}_\mathrm{lmfp}^{-1}+\hat{\kappa}_\mathrm{smfp}^{-1})$. For the relative velocity dependent cross section scenarios the time interval mapping relation is: 
\begin{equation}\label{eq:kappa_general}
    \delta\hat{t}_b=\dfrac{\langle\hat{\kappa}_av_{12}^2\rangle_{\hat{t}_a}}{\langle\hat{\kappa}_bv_{12}^2\rangle_{\hat{t}_a}}\delta\hat{t}_a\,,
\end{equation}
here the $v_{12}$ terms in the MB integration comes again from the energy transfer. The challenge of computing Eq~\ref{eq:kappa_general} compared to the asymptotic lmfp and smfp scenarios is that constant parameters $\beta$, $\hat{\sigma}_c$, as well as the scale-free density $\hat{\rho}$ can no longer be pulled out from the integration due to this more complicated heat conductivity expression. It is therefore necessary to prepare a 4D lookup table that simultaneously covers the variations of $\{\beta, \hat{\sigma}_c, \hat{\omega}/\hat{v}_\mathrm{rms},\hat{\rho}\}$ to achieve fast mapping calculations.\par

\section{Observational constraints on the cross section model}\label{sec:pConstrain}
Both the gravothermal fluid formalism and idealized N-Body simulations are computationally expensive in the sense that they cannot be directly used to explore the continuous SIDM particle parameter space. The strength of the lmfp mapping method introduced in this work is that one needs to numerically solve for the gravothermal fluid equations only once, and can then extend this solution to halos with arbitrary cross section models, as long as the halo is mostly evolving in the lmfp regime. It therefore serves as a powerful tool for deriving constraints on SIDM cross section model parameters.\par
Halos hosting dwarfs and LSB galaxies are promising laboratories for constraining SIDM models because these small galaxies serve as powerful tracers of the DM halo density profile, while their shallow baryonic potentials leave little influence on the DM dynamics (although see \citealt{2012MNRAS.421.3464P,2016MNRAS.459.2573R} for example). As a demonstration of the utility of this work, we constrain the cross section model parameter space favored by the rotation curve measurement of a LSB dwarf galaxy UGC 128 \citep{1993AJ....106..548V,2017PhRvL.119k1102K}. We select UGC 128 because in this system the circular velocity is mostly determined by the dark matter component, while the stellar disc has negligible effect \citep{2017PhRvL.119k1102K}. The fact that our model does not capture baryonic physics is therefore not expected to significantly influence the SIDM parameters inferred from UGC 128's rotation curve. We notice that the rotation curves of dozens of dwarf/LSB galaxies have been measured carefully \citep[e.g.][]{2008ApJ...676..920K,2015AJ....149..180O,Essig2019}. Some of the measurements have been used to derive stringent upper bounds on cross section models \cite[e.g.][]{2018MNRAS.481..860R,2021arXiv210803243J}. While a careful and thorough exploration of the SIDM cross section model parameter space using all the available dwarf/LSB observational data will be an important step to take, in this work we choose to show just an example application of the gravothermal solution mapping method. We therefore illustrate parameter space regions favored by the data on UGC 128 alone and defer a more complete analysis of a larger dwarf/LSB galaxy sample to a subsequent work. In this work we trust the UGC 128 rotation curve measurement and ignore potential error underestimation caused by beam smearing, coherent turbulence, the
lack of determination of pressure support, and the treatment of inclination
errors \citep{2019MNRAS.482..821O,2021MNRAS.502.3843S}. The derived constraints can therefore be over-tightened.\par
\begin{figure*}[t!]
    \centering
    \includegraphics[width=0.49\textwidth]{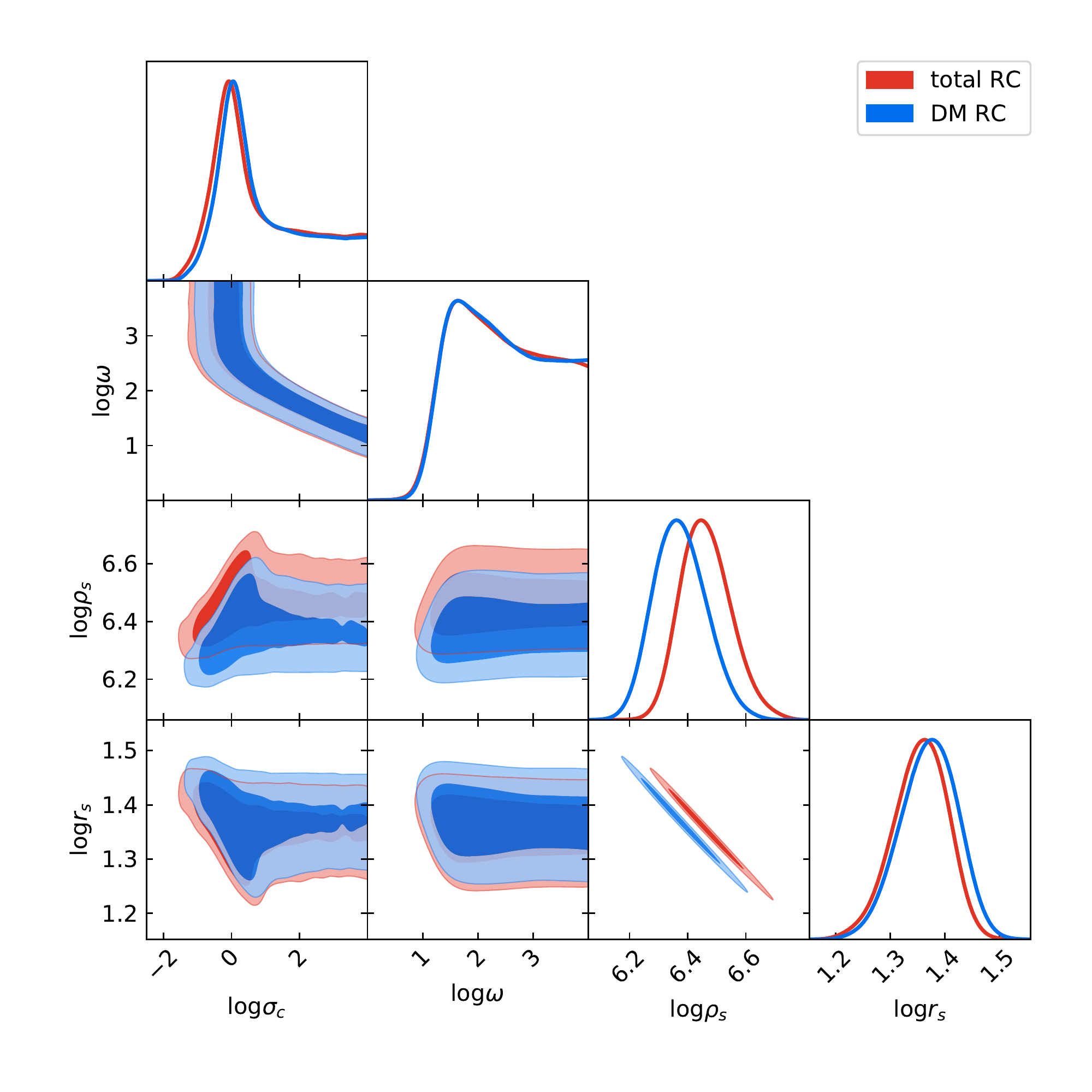}
    \includegraphics[width=0.49\textwidth]{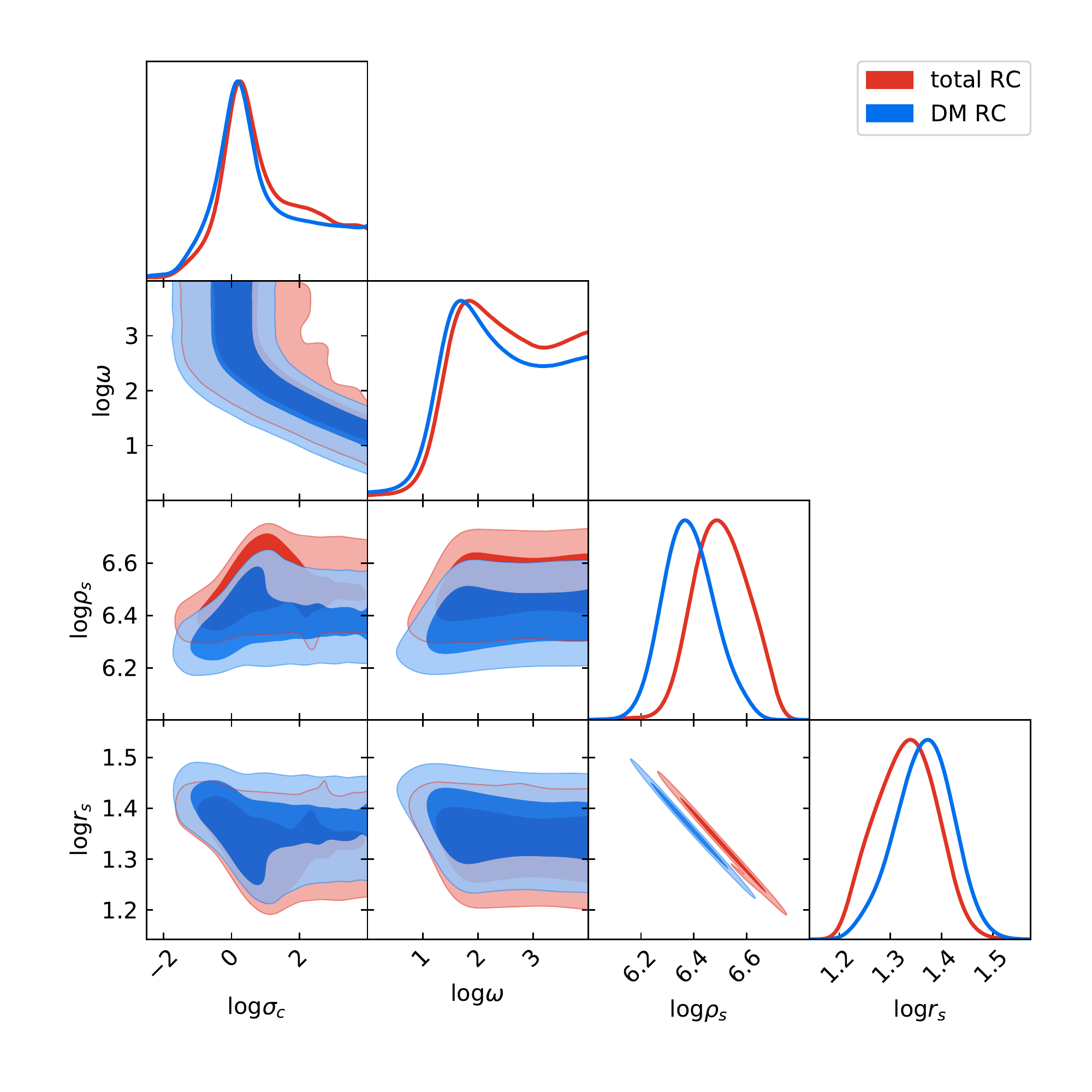}
    \vspace{-5mm}
    \caption{Parameter constraints for the NFW initial condition Eq~\ref{eq:NFW} and the cross section model  Eq~\ref{eq:Gilman2021} resulting from dwarf galaxy UGC 128 \citep{1993AJ....106..548V,2017PhRvL.119k1102K} rotation curve measurements. We fix $\beta=0.5$ and a halo evolution time of 10 Gyr. We assume Gaussian likelihood for the total/DM only rotation curve measurements and show the parameter space favored by observation in red/blue regions. The contours enclose parameter constraints at 68\% and 95\% confidence levels. The left panel shows MCMC fitting results derived from the original UGC 128 rotation curve, while for the right panel the rotation curve error bars for the innermost three radius bins ($r\leq 5$ kpc) are enlarged by a factor of 2 to account for potential measurement uncertainty underestimation.}
    \label{fig:pConstrain}
\end{figure*}

We assume $\beta=0.5$ and a halo evolution time of 10 Gyr. We then use \textsc{emcee} \citep{2013PASP..125..306F} to conduct the \textsc{MCMC} sampling over the 4D parameter space $-2.5\leq\log\sigma_c/[\mathrm{cm^2/g}]\leq4.0$, $0.0\leq\log\omega/[\mathrm{km/s}]\leq4.0$, $5.0\leq\log\rho_s/[M_\odot/\mathrm{kpc}^3]\leq8.0$, and $-1.0\leq\log r_s/[\mathrm{kpc}]\leq 2.0$, assuming uniform priors on these logarithmic variables. Here $\sigma_c$ and $\omega$ are parameters introduced by the velocity dependent cross section model $\sigma(v_{12})=\sigma_c/(1+v_{12}^2/\omega^2)^2$, and $\{\rho_s,r_s\}$ are the NFW profile scaling parameters. In each \textsc{MCMC} step we use the gravothermal mapping method introduced in Section~\ref{sec:matchArepo} to compute the halo density profile at $t=10$ Gyr, and transfer the dark matter density profile to rotation curve through $v_\mathrm{circ}(r)=\sqrt{GM(<r)/r}$. Here $v_\mathrm{circ}(r)$ is the circular velocity at radius $r$. We adopt a Gaussian likelihood function to direct 20 \textsc{MCMC} random walkers with the UGC 128 rotation curve measurement. To get a quantitative sense of how strongly the baryonic gravitational potential in this system may influence the parameter fitting results, we perform two independent \textsc{MCMC} fits for the total rotation curve and the dark matter halo rotation curve, respectively. Contributions to the rotation curve from stars and gas are fit by \cite{2017PhRvL.119k1102K}. A triangle plot for the parameter constraints is presented in Figure~\ref{fig:pConstrain} left panel.\par 
The \textsc{MCMC} fitting results show that the NFW parameters $\rho_s$ and $r_s$ are more sensitive to the differences between the total and dark matter only rotation curves, while the $\log\sigma_c-\log r_s$ parameter constraints show little response to this data variation. This is fortunate as we are generally more interested in learning about the cross section parameters rather than fitting NFW parameters of different halos with high accuracy. However, the generality of this feature should be further tested using a larger observational sample.\par 
The \textsc{MCMC} results also show that the NFW scaling parameters $\rho_s$ and $r_s$ are degenerate. This is expected since the density profile amplitude increase caused by increasing $\rho_s$ can be compensated by decreasing $r_s$, which effectively shifts the halo density profile to smaller radii. Despite this parameter degeneracy we are still able to constrain $\{\rho_s,r_s\}$ within a range much narrower than the flat priors. The best-fit NFW parameters for this halo fitted from the total rotation curve are $\log\rho_s/[M_\odot/\mathrm{kpc}^3]=6.49^{+0.17}_{-0.16}$ and $\log r_s/[\mathrm{kpc}]=1.34\pm0.09$ ($1\sigma$ errors). For the rotation curve with the fitted stellar disc and gas components subtracted, the best-fit NFW scales are $\log\rho_s/[M_\odot/\mathrm{kpc}^3]=6.40\pm0.17$ and $\log r_s/[\mathrm{kpc}]=1.35\pm0.10$. There is a ``L"-shaped degeneracy between the cross section parameters $\log\sigma_c$ and $\log\omega$. One can also see such a degeneracy from the asymptotic behavior of the cross section model. Specifically, the cross section model reduces to the constant factor $\sigma_\mathrm{c}$ for high $\omega$, and the parameter constraints become independent of $\omega$. On the other hand, lowering $\omega$ effectively decreases the cross section, resulting in a weaker upper bound on $\sigma_\mathrm{c}$. The turn over of this ``L''-shaped band occurs on a scale comparable to the characteristic rms velocity of the halo. Therefore, the degeneracy between $\sigma_c$ and $\omega$ can be at least partially broken by combining similar parameter constraints for systems of different sizes, masses, and characteristic velocity dispersions, and we defer a more careful analysis to a future work. We also emphasize that in this fit we have assumed $\beta=0.5$ and halo evolution time of 10 Gyr. Due to the $\beta\hat{\sigma}\hat{t}$ degeneracy introduced in Sections~\ref{sec:mappingEssig}-\ref{sec:matchArepo}, assuming alternative $\beta$ and halo evolution time values would rescale the $\sigma_c$ constraints by a factor of $1/(\beta t)$. The constraints on parameters $\{\omega,\rho_s,r_s\}$ are insensitive to the $\beta$ and halo age assumptions.\par
\begin{figure*}
    \centering
    \includegraphics[width=0.9\textwidth]{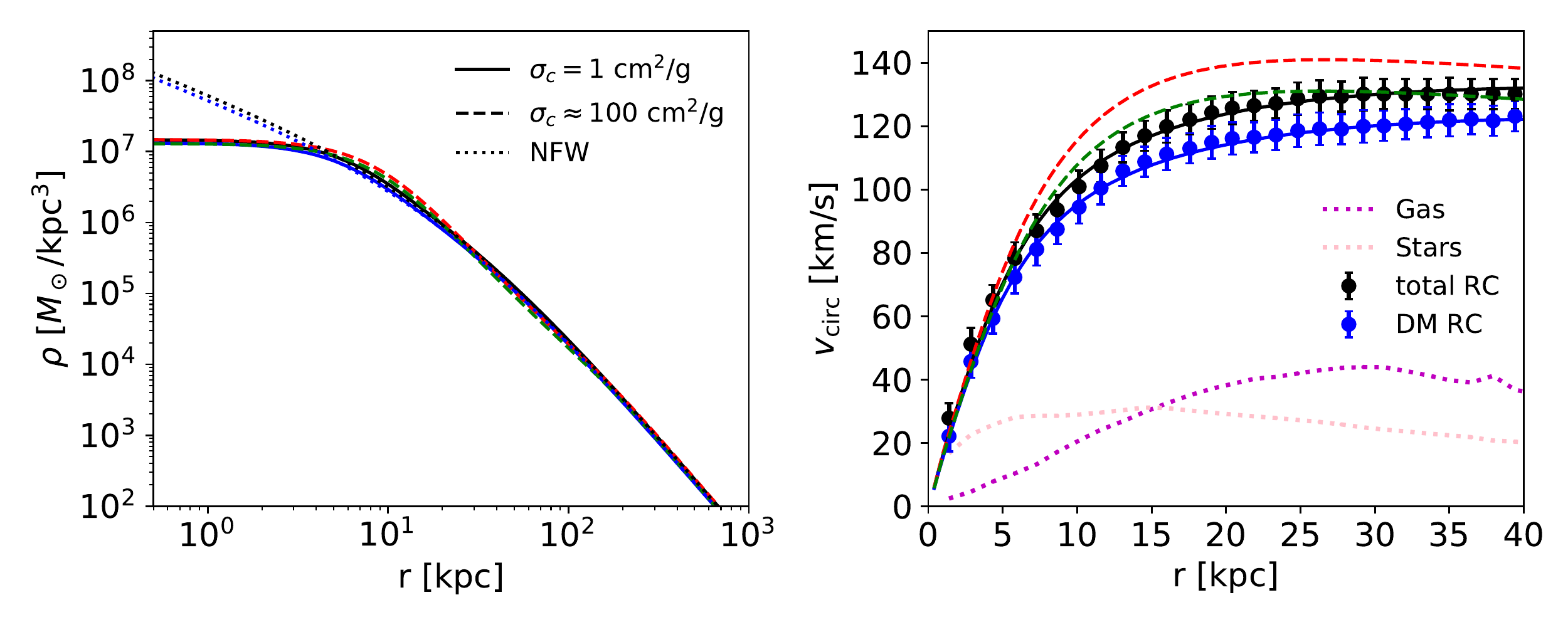}
    \caption{The best-fit halo density profiles (left) and rotation curves (right) at 10 Gyr, assuming $\beta=0.5$ but different SIDM cross sections. The black/blue solid curves show the best-fit mapped gravothermal solution to the total/DM only rotation curve. These two best-fit solutions correspond to a small cross section $\sigma_c=1$ cm$^2$/g, $\omega=10^4$ km/s.The red/green dashed lines show mapped gravothermal solutions under a larger cross section $\sigma_c \approx 100$ cm$^2$/g, $\omega=10^4$ km/s, but identical halo central densities to the total/DM only rotation curve best-fit solutions. Compared to the NFW profiles with the total/DM only rotation curve best-fit $\{\rho_s, r_s\}$ parameters shown by the black/blue dotted curves, density profiles for the $\sigma_c = 1$ cm$^2$/g case is only suppressed in the cored region, while a larger region in the halo is influenced by the DM self-interaction for the $\sigma_c \approx 100$ cm$^2$/g case. Only the small cross section solutions match the observed total/DM only rotation curves, shown by black/blue data points. The purple and pink dotted curves show rotation curves contributed by gas and stars in UGC 128, fitted by \cite{2017PhRvL.119k1102K}.}
    \label{fig:twoSolutions}
\end{figure*}

The single ``L"-shaped band in the $\log\sigma_c-\log\omega$ parameter space favored by the observed rotation curve was a surprise to us. Specifically, we have assumed an NFW initial density profile, for which the halo central density first decreases, reaching a maximum core size, and then increases with time. There will be in general two $\beta\sigma t$ instances where the cored halo density profiles are very similar to each other (unless the halo is measured to be precisely at the moment of maximum core size, or the halo has entered the late core collapsing stage where its central density is higher than the initial state): one from the core formation process and another one from the core collapsing process. The existence of two gravothermal solutions with similar halo density profiles but very different $\beta\hat{\sigma}\hat{t}$ is one of the causes of the huge SIDM cross section constraint dispersion among different works \citep{2020JHEP...12..202K}. For example, SIDM cross sections of Milky Way's dwarf satellite galaxies constrained by isothermal models such as \cite{2017PhRvL.119k1102K} and \cite{2018NatAs...2..907V} only pick out the low cross section solution because the isothermal models by design only works for the SIDM halo core formation process. Cross sections constrained by isothermal-based methods are therefore relatively low, ranging from 0.1 to 40 cm$^2$/g. On the other hand, previous studies such as \cite{2021MNRAS.503..920C} that use the gravothermal fluid formalism to constrain SIDM cross sections may only focus on the halo core collapsing process, and give higher constraints 30--200 cm$^2$/g for similar systems. We find the ``L''-shaped band favored by the UGC 128 rotation curve points to one solution during the core formation process. As an example, we pick one point in the best-fit parameter space $\sigma_c=1$ cm$^2/g$, $\omega=10^4$ km/s, and compare the mapped halo density profile as well as the rotation curve at $t=10$ Gyr with observations in Figure~\ref{fig:twoSolutions}. We show the best-fit gravothermal solutions to the total and DM only rotation curves in black and blue solid curves, respectively. We find fixing the best-fit $\rho_s$, $r_s$, and $\omega$ values, but enlarging $\sigma_c$ to about 100 cm$^2$/g gives very similar mapped halo density profile at $t = 10$ Gyr, although under such a large cross section the halo has started core collapse. The gravothermal solutions with identical halo central density to the best-fit solution, but a much larger cross section is shown in dashed lines in Figure~\ref{fig:twoSolutions}. The red/green dashed line corresponds to the total/DM only rotation curve case. We show in the right panel of Figure~\ref{fig:twoSolutions} that, for either the total or the DM only case, rotation curve measurements of UGC 128 at radii less than about 5 kpc do not favor either one of those two gravothermal solutions, owning to their similar halo central densities. To clarify the origin of this tight cross section constraint from UGC 128, in the left panel of  Figure~\ref{fig:twoSolutions} we compare these two sets of gravothermal solutions to NFW profiles with the total/DM-only rotation curve best-fit $\{\rho_s, r_s\}$ parameters, shown in black/blue dotted curve. During the halo core formation process where $\sigma_c = 1$ cm$^2$/g, the halo density profile is only suppressed in the cored region ($r\lesssim5$ kpc) compared to the NFW initial condition. However, for the $\sigma_c \approx 100$ cm$^2$/g case the halo density is altered by DM self-interactions over a larger radial range ($r\lesssim100$ kpc). As a result the halo density profile outside the cored region becomes steeper than the initial NFW profile. This subtle difference is distinguishable by the rotation curve measurements outside the isothermal core. However, we emphasize that all the above analysis relies on the assumption that the UGC 128 rotation curve data and its measurements uncertainties are statistically valid. The rotation curve error bars can be underestimated, especially at the galaxy central region, due to center offsets, non-circular motions, inclination, the effects of pressure support, and many other observational uncertainties (see \citealt{2010AdAst2010E...5D} for a review). While a careful reprocessing of the UGC 128 rotation curve data is beyond the scope of this work, we attempt to account for potentially underestimated uncertainties through increasing the UGC 128 rotation curve error bars of the innermost three radius bins, where the radius is less than 5 kpc, by a factor of 2. The MCMC fitting results are shown in Figure~\ref{fig:pConstrain} right panel. Assuming larger measurement uncertainties near the galaxy center, the ``L"-shaped bands in the $\log\sigma_c-\log\omega$ space become significantly wider such that cross sections as large as $\sigma_c\sim100$ cm$^2$/g can be preferred by the rotation curve measurement at 95\% confidence level.

\par

We note that dissipative cooling processes \citep{Essig2019}, the effects of a baryonic disc \citep{2022arXiv220612425J}, and tidal effects \citep{2022MNRAS.513.4845Z} can all speed up the onset of halo core collapse. On the other hand, \cite{2022arXiv221001817M} suggests that halo mass accretion and merger can delay the core collapse process, and proposes a correction term to the halo core collapsing time in order to account for this effect. However, the goal of this work is simply to introduce a convenient non-linear time rescaling method for the gravothermal fluid formalism without considering any of the above effects. 
Nevertheless, such a preliminary constraint already shows the advantage of our mapping method in the sense that it can rapidly explore the continuous parameter space, and clearly present the parameter degeneracies.\par
\par

\section{Conclusion and discussion}\label{sec:concluson}
In this work we introduce a fast mapping method that transfers the gravothermal fluid formalism solutions from the constant cross section scenario to arbitrary velocity-dependent cross section models. Such a mapping is intrinsically built in the gravothermal fluid formalism. Specifically, when a halo is evolved fully in either the lmfp or the smfp regime, cross section only influences the heat conductivity of the isolated halo at each radius and moment, which further determines the halo evolution rate. Varying the cross section model assumptions therefore only re-maps the time axis of the halo evolution, but has no effect on the radial profile amplitudes/shapes of halo density, velocity, enclosed mass, etc. Such a time re-mapping is linear between two different constant cross sections, but becomes non-linear between constant and velocity dependent cross section models.\par 
The thermal conductivity of a SIDM halo varies at different radii and times. To achieve fast mapping we use the thermal conductivity at the halo center at each instance as a characteristic value. We test the performance of this mapping method through comparing the mapped gravothermal solutions, assuming a constant cross section, with actual gravothermal numerical solutions assuming different velocity dependent cross section models. We find that the fast mapping method introduced in this work nicely captures the full time evolution of a SIDM halo, although it slightly overestimates the core collapsing time due to the simplified assumption about the halo instantaneous heat conductivity. We show that the accuracy of this mapping method can be further boosted through a more careful treatment of the instantaneous heat conductivity calculation, but argue that since the fractional error of the default mapping is small compared to the uncertainties of halo evolution time and the free parameter $\beta$, the more complicated mapping method is unnecessary.\par 
Calibrating gravothermal solutions with N-Body simulations under the assumption of a velocity dependent cross section is more challenging since in numerical simulations the particle scattering probabilities are generally determined by their relative velocities, while the constant cross section gravothermal fluid formalism only cares about the particle rms velocity/velocity dispersion in each radial shell. Assuming the SIDM particle velocity in each halo shell follows the Maxwell-Boltzman distribution, we define the averaged thermal conductivity as $\langle\kappa_\mathrm{lmfp}\rangle\propto\langle v_1v_2v_{12}^3\sigma(v_{12})\rangle$, where $v_1$, $v_2$ are the velocities of two particles, and $v_{12}$ is the relative velocity. The velocity terms in the MB-distribution averaging bracket are contributed by particle orbiting scales, collisional relaxation times, and energy transfer. We show that this physically motivated definition of the averaged heat conductivity provides an excellent match between the mapped gravothermal solutions and \textsc{Arepo} idealized N-Body simulation results, assuming multiple different velocity dependent cross section models.\par
The advantage of this mapping method lies in its high computational efficiency. Specifically, it takes about 24 CPU hours of computation with our gravothermal fluid code to reach the core-collapse regime for an SIDM halo in the lmfp regime, and introducing a velocity dependent cross section model brings at least two extra degree-of-freedom. It is therefore impractical to continuously explore the cross section model parameter space directly with the gravothermal fluid code. Luckily, by adopting the mapping method introduced in this work one only needs to solve the fluid code for once, and can fast map this set of numerical solutions to halos with arbitrary cross section assumptions. Besides deriving parameter constraints, this mapping method is also suitable to be implemented into semi-analytic galaxy formation models in order to achieve fast SIDM halo population generation.\par
As an example application, we select the LSB system UGC 128 where the baryonic disc potential has negligible influence on the galaxy rotation curve, and use the mapping method to constrain 4 free parameters in the NFW and velocity-dependent cross section models. We have assumed $\beta=0.5$ and a halo evolution time of 10 Gyr, but we show analytically that the $\beta t$ assumption will only re-scale constraints of the cross section scaling parameter $\sigma_\mathrm{c}$. The other three parameters $\{\omega,\rho_\mathrm{s},r_\mathrm{s}\}$ are not sensitive to the assumed $\beta$ and halo age values. To get a quantitative sense of how much the baryonic gravitational potential in this system may influence the parameter fitting results, we perform two independent \textsc{MCMC} fits for the total rotation curve and the dark matter halo rotation curve, respectively, finding that the SIDM parameters of interest are unaffected. We also show that the accurately measured UGC128 rotation curve can distinguish two very similar gravothermal solutions during the core formation and core collapsing phases, and favors the low cross section solution. We notice that the constraints derived in this work do not account for other possible non-linear effects that can influence the halo core collapsing time, including cooling, an additional gravitational potential contributed by a baryonic disc or black hole, tidal effects, halo mass accretion, and merger. The constraints can be too tight since we have trusted the UGC128 rotation curve measurements, and have ignored possible error underestimation. Nevertheless the advantage of this fast gravothermal solution mapping method is recognized in the sense that it can quickly explore the continuous 4D parameter space with the first-principle gravothermal solutions, and effectively capture degeneracies among the SIDM halo parameters of interest.\par
We also compare the performance of this mapping method with a similar strategy proposed in \citetalias{2022arXiv220406568O}. We find the simpler method introduced in \citetalias{2022arXiv220406568O}, where the characteristic heat conductivity is assumed to be a time and radius independent constant, outperforms the method introduced in this work in reproducing the core collapse time when mapping between gravothermal solutions with $v_\mathrm{rms}$ dependent cross section models. However, this work performs better in mapping the full halo evolution process, while \citetalias{2022arXiv220406568O} fails to capture the halo central density evolution before the instant of maximum core size if the cross section model has strong velocity dependency. Correctly mapping the SIDM halo core formation process is as important as capturing the core collapse time because: 1) most current SIDM cosmological simulations evolve halos with mild self-interaction cross section for only around 10~Gyr, such that the simulation stops before an isothermal core has been completely formed (e.g.\ \citealt{2015MNRAS.453...29E,2020ApJ...896..112N}), and 2) observations may prefer a SIDM model with small cross section. If this is the case, most halos hosting observable galaxy clusters/galaxies/dwarfs are still approaching their maximum core size. \citetalias{2022arXiv220406568O} propose to estimate the averaged thermal conductivity in a particle relative velocity dependent scenario as $\langle\kappa_\mathrm{lmfp}\rangle\propto\langle v_{12}^n\sigma(v_{12})\rangle$, where $n$ is to be calibrated to N-Body simulations. We find that a physically motivated definition $\langle\kappa_\mathrm{lmfp}\rangle\propto\langle v_1v_2v_{12}^3\sigma(v_{12})\rangle$ outperforms the ansatz assumed in \citetalias{2022arXiv220406568O}, where $n=3$. However, we show that combining the linear time re-scaling method introduced by \citetalias{2022arXiv220406568O} and the definition $\langle\kappa_\mathrm{lmfp}\rangle\propto\langle v_{12}^5\sigma(v_{12})\rangle$, also suggested by \cite{2022JCAP...09..077Y}, gives an equally good match to the N-Body simulation results as $\langle\kappa_\mathrm{lmfp}\rangle\propto\langle v_1v_2v_{12}^3\sigma(v_{12})\rangle$.\par

\section{Acknowledgements}
We thank Yiming Zhong for sharing his gravothermal fluid code, and for many helpful discussions. We thank Mark Vogelsberger for sharing the \textsc{Arepo} N-body simulation package including the SIDM module. Computing resources used in this work were made available by a generous grant from the Ahmanson Foundation. This work was supported in part by the NASA Astrophysics Theory Program under grant 80NSSC18K1014.\par

\appendix
\section{An empirical model for the lmfp gravothermal density profiles.}\label{apx:empiricalModel}
In this appendix we introduce a convenient empirical model for the density profile of SIDM halos, calibrated to the lmfp gravothermal solution used in this work.\par
\cite{2002ApJ...568..475B} has proved that the gravothermal density profile evolves with time self-similarly, and the density in the halo outskirts falls with radius as $\hat{\rho}\propto\hat{r}^{-2.19}$. This derivation is elegant, but not directly applicable to most of the numerical SIDM simulations. Specifically, in most idealized N-Body as well as some cosmological simulations the halo starts from an NFW initial state where $\hat{\rho}\propto\hat{r}^{-3}$ at large radii. As a result, when a core has been formed the halo density still drops as $\hat{\rho}\propto\hat{r}^{-2.19}$ slightly outside the core radius. At even larger radii the density profile joins smoothly with the NFW initial condition, and drops as $\hat{\rho}\propto\hat{r}^{-3}$. At this stage the halo density profile can be captured by a triple power law.\par
We first define the instant of halo core collapse as the time at which the halo central density diverges:
\begin{equation}
    \log(\beta\hat{\sigma}\hat{t})=E=2.238\,,
\end{equation}
and the instant of maximum core size as the time when the halo central density reaches its minimum:
\begin{equation}
    \log(\beta\hat{\sigma}\hat{t})=F=1.341\,.
\end{equation}
We find at the early stage of the halo evolution, i.e. $\log(\beta\hat{\sigma}\hat{t})< F$, the halo density profile can still be described by the NFW profile multiplied by a tanh central cut:
\begin{equation}
    \hat{\rho}=\dfrac{\tanh(\hat{r}/r_\mathrm{core})}{\hat{r}(1+\hat{r})^2}\,,
\end{equation}
where we use free parameter $r_\mathrm{core}$ to capture the core size. We find the following empirical model described the time evolution of the halo core size at $\log(\beta\hat{\sigma}\hat{t})\leq1.341$ with high accuracy:
\begin{equation}\label{eq:ro_pro}
        \log r_\mathrm{core}=A_{r_\mathrm{core}}(\log(\beta\hat{\sigma}\hat{t}))^2+B_{r_\mathrm{core}}\log(\beta\hat{\sigma}\hat{t})+C_{r_\mathrm{core}}\,.
\end{equation}
Here $A_{r_\mathrm{core}}=-0.1078$, $B_{r_\mathrm{core}}=0.3737$, $C_{r_\mathrm{core}}=-0.7720$ are calibrated to the gravothermal solutions.\par
As mentioned before, in later phases of halo evolution, $\log(\beta\hat{\sigma}\hat{t})>1.341$, when a core has been formed, the density profile can be described by triple power law:
\begin{equation}
    \hat{\rho}(\hat{r})=\dfrac{\rho_\mathrm{core}}{1+\left(\dfrac{\hat{r}}{r_\mathrm{core}}\right)^{s}\left(1+\dfrac{\hat{r}}{r_\mathrm{out}}\right)^{3-s}}\,.
\end{equation}
Here $s=2.19$ is provided by \cite{2002ApJ...568..475B} gravothermal solution self-similar analysis. Again we use free parameter $r_\mathrm{core}$ to trace the characteristic halo core size at each time step. Similarly, the parameter $\rho_\mathrm{core}$ captures the halo central density, and $r_\mathrm{out}$ traces the radius where the density profile influenced by self-interaction joins smoothly to the NFW outskirts. We develop the following empirical model for $\rho_\mathrm{core}$, $r_\mathrm{core}$, and $r_\mathrm{out}$, calibrated to the lmfp gravothermal solution:
\begin{equation}
    \log\rho_\mathrm{core}=A_{\rho_\mathrm{core}}(\log\beta\hat{\sigma}\hat{t}-(E+3))^2+C_{\rho_\mathrm{core}}+\dfrac{D_{\rho_\mathrm{core}}}{(E+0.0001-\log\beta\hat{\sigma}\hat{t})^{0.02}}\,,
\end{equation}
where $A_{\rho_\mathrm{core}}=0.05771,\ C_{\rho_\mathrm{core}}=-21.64,\ D_{\rho_\mathrm{core}}=21.11$,\par
\begin{equation}
    \log r_\mathrm{core}=A_{r_\mathrm{core}}(\log\beta\hat{\sigma}\hat{t}-(E+2))^2+C_{r_\mathrm{core}}+\dfrac{D_{r_\mathrm{core}}}{(E+0.0001-\log\beta\hat{\sigma}\hat{t})^{0.005}}\,,
\end{equation}
where $A_{r_\mathrm{core}}=-0.04049,\ C_{r_\mathrm{core}}=43.07,\ D_{r_\mathrm{core}}=-43.07$, and\par
\begin{equation}\label{eq:ro_pos}
    \log r_\mathrm{out}=A_{r_\mathrm{out}}(\log\beta\hat{\sigma}\hat{t}-(E+2))^2+C_{r_\mathrm{out}}+\dfrac{D_{r_\mathrm{out}}}{(E+0.04-\log\beta\hat{\sigma}\hat{t})^{0.005}}\,,
\end{equation}
where $A_{r_\mathrm{out}}=0.02403,\ C_{r_\mathrm{out}}=-4.724,\ D_{r_\mathrm{out}}=5.011$.\par
\begin{figure}[t!]
    \centering
    \includegraphics[width=0.5\textwidth]{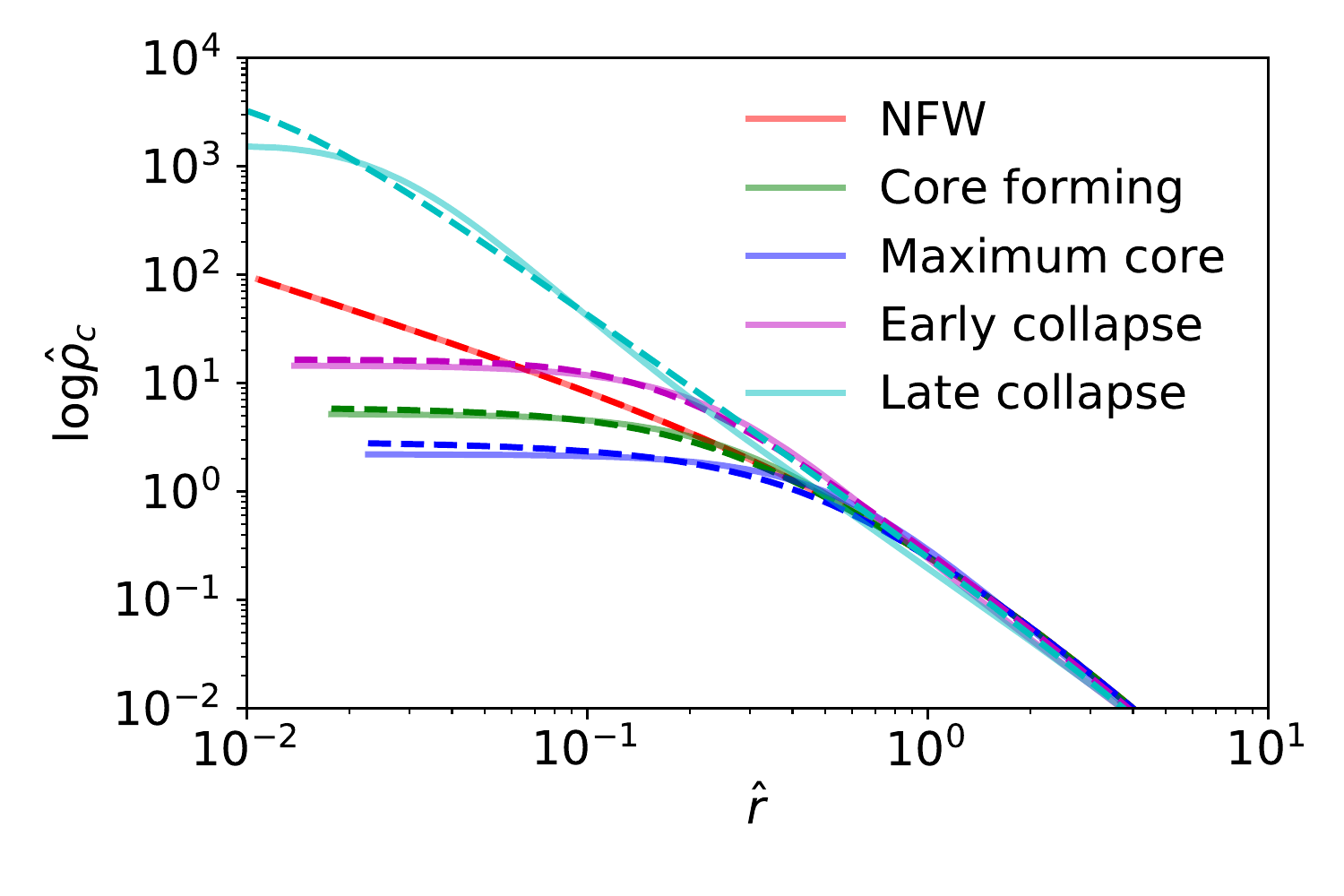}
    \caption{Density profile comparisons between the gravothermal solutions and the empirical model introduced in this section. We compare the numerical solution and empirical model at the halo initial state $\beta\hat{\sigma}\hat{t}=0$ (red), core formation state $\beta\hat{\sigma}\hat{t}<F=1.341$ (green), maximum core moment $\beta\hat{\sigma}\hat{t}=F$ (blue), early core collapse $\beta\hat{\sigma}\hat{t}=150$ (magenta), and the late core collapse state $\beta\hat{\sigma}\hat{t}=E=2.238$ (cyan). The gravothermal solution density profiles are shown as faint solid curves, while the empirical model predictions are shown in the dashed lines.}
    \label{fig:empirical_rhor}
\end{figure}
\begin{figure}
    \centering
    \includegraphics[width=0.45\textwidth]{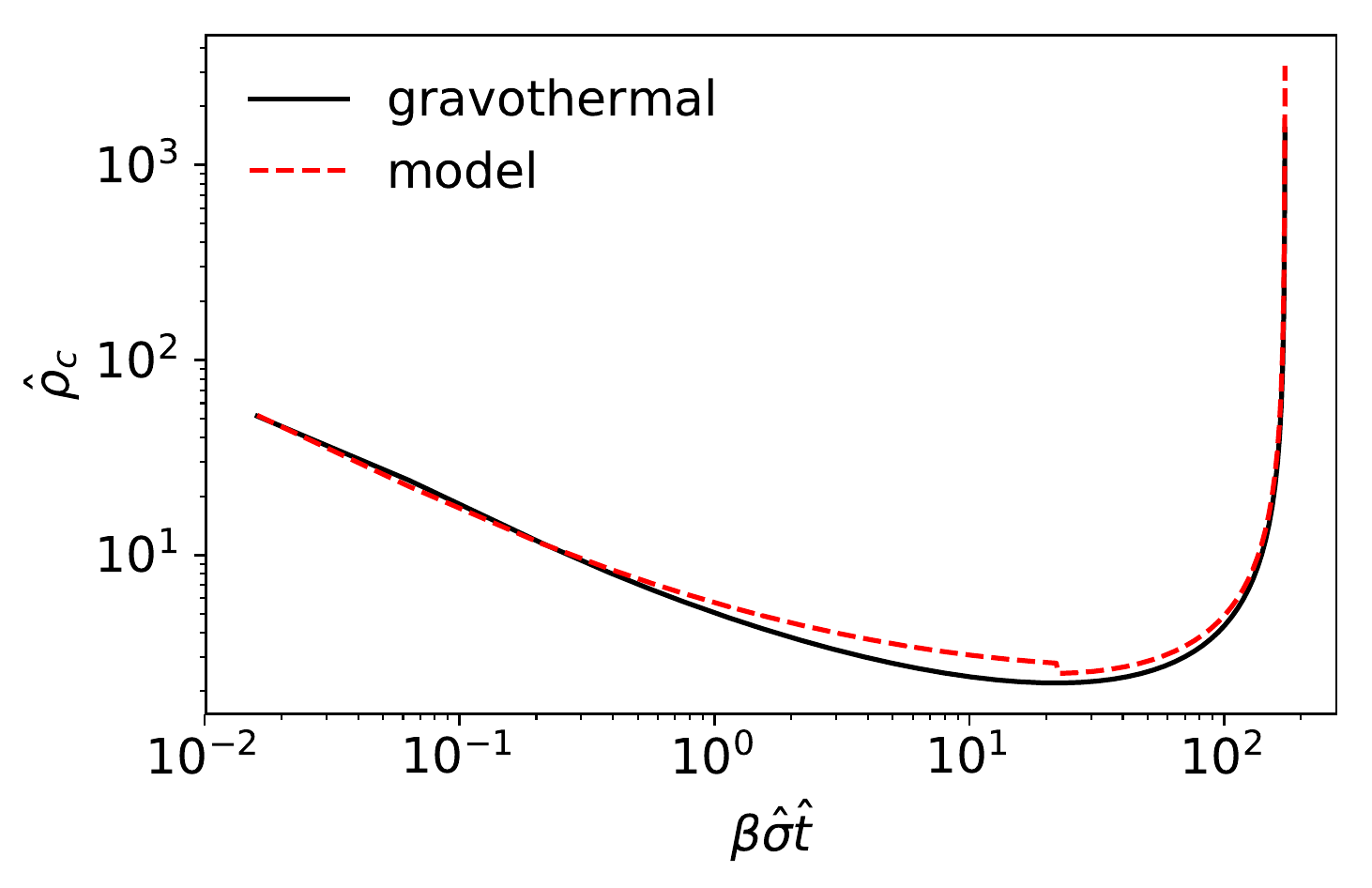}
    \includegraphics[width=0.45\textwidth]{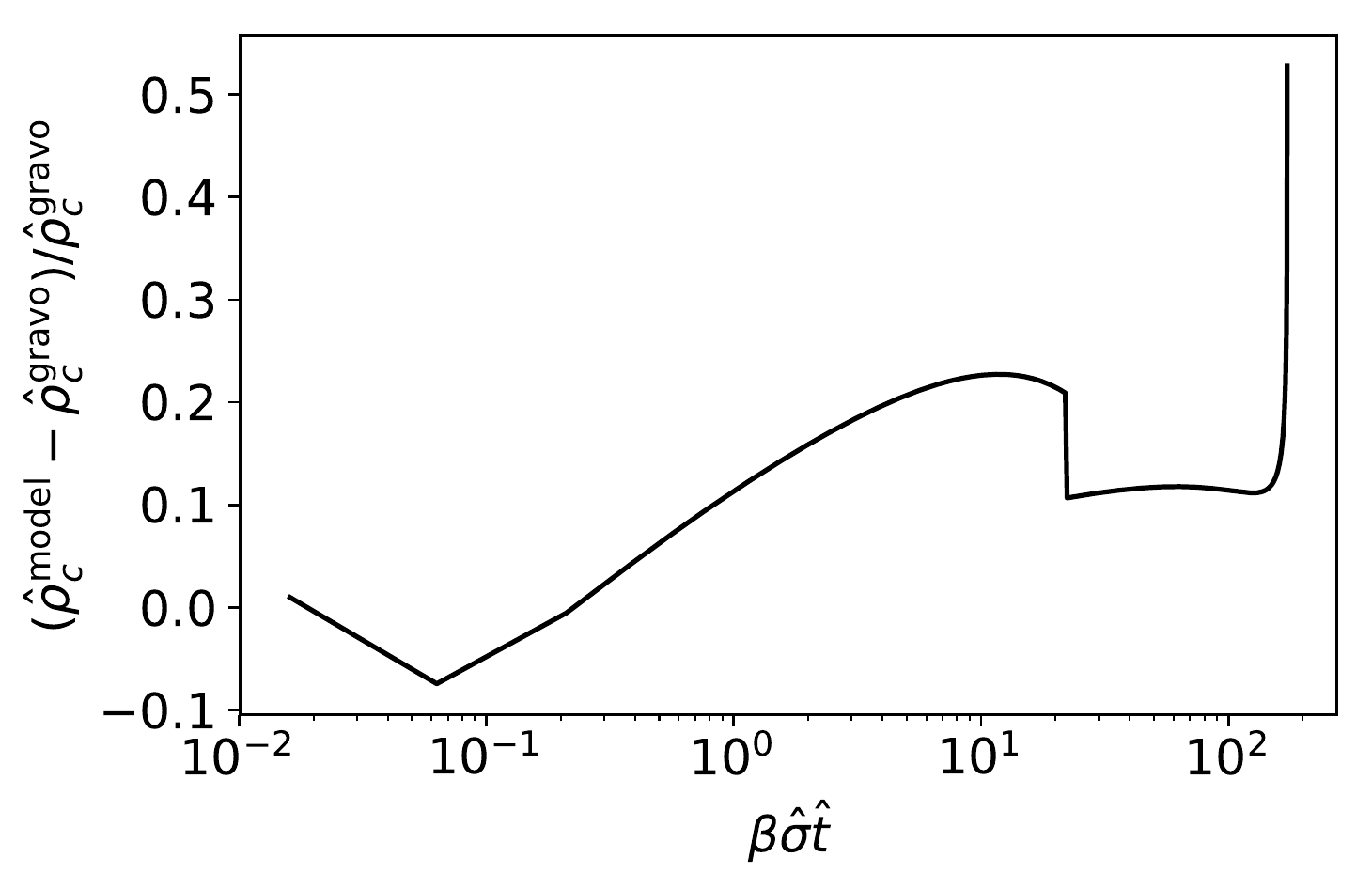}
    \caption{\textit{Left:} Halo central density time evolution comparisons between the gravothermal solutions (black solid curve) and empirical model (red dashed curve) introduced in this section. \textit{Right:} Fractional error of the empirical model in reproducing the halo central density given by the gravothermal solutions.}
    \label{fig:empirical_rhoc}
\end{figure}

We compare the halo density profile given by the gravothermal solution and this empirical model at five characteristic moments in Figure~\ref{fig:empirical_rhor}. This simple empirical model accurately captures the overall time evolution and radial trend of the gravothermal solutions throughout the halo evolution process. We also show the halo central density using the exact result from the gravothermal fluid model and our empirical model (along with their fractional difference) in Figure~\ref{fig:empirical_rhoc}. The fractional error of this empirical model is better than 10\% at most times. Even at the very late core collapsing stage its accuracy is still better than 50\%. This degree of accuracy is sufficient for many applications, considering the fact that the random fluctuations in N-Body simulation results can cause the halo central density to fluctuate at a similar level. The fractional error shows non-continuous behavior at the maximum core moment $\beta\hat{\sigma}\hat{t}=1.341$ because we switch the density profile empirical description from a tanh cutoff to the triple power law here.\par 
This analytic empirical model has wide applications. For example, it can be implemented into semi-analytic galaxy formation models for fast SIDM halo simulation. As another example, we show in Section~\ref{sec:mapping} that the characteristic radius $r_\mathrm{out}$ provided by this model can be used to boost the accuracy of the mapping method introduced in this work.\par

\bibliography{gravothermalMapping}{} 
\bibliographystyle{aasjournal}

\end{document}